\documentclass[prl,aps,nofootinbib,twocolumn,noshowpacs,noshowkeys,superscriptaddress]{revtex4-2}
\usepackage{comment}

\usepackage{hyperref}
\hypersetup{colorlinks=true, citecolor=[rgb]{0,0,0.8}, urlcolor=[rgb]{0,0,0.8}, linkcolor=[rgb]{0,0,0.8}}

\usepackage{amsfonts}
\usepackage{amsmath}
\usepackage{amssymb}
\usepackage{graphicx}

\usepackage{textcomp}
\newcommand{\vv}[1]{\mathbf{#1}}

\usepackage{color}

\begin{document}

\def\neel{{Institut N\'{e}el, Universit\'{e} Grenoble Alpes - CNRS:UPR2940, 38042 Grenoble, France}\\

$^*$ These authors contributed equally to this work }

\author{Clement Gouriou$^*$}
\affiliation{\neel}
\author{Cattleya Dousset$^*$ }
\affiliation{\neel}
\author{Alex Fontana}
\affiliation{\neel}
\author{Antoine Reigue}
\affiliation{\neel}
\author{Francesco Fogliano}
\affiliation{\neel}
\author{Hugo Weltz}
\affiliation{\neel}
\author{Lucas Jud{\'e}aux}
\affiliation{\neel}
\author{Michael Croquette}
\affiliation{\neel}
\author{Benjamin Pigeau}
\affiliation{\neel}
\author{Olivier Arcizet}
\affiliation{\neel}
\email{olivier.arcizet@neel.cnrs.fr}

\title{
Nano-optomechanical exploration of the dynamical photothermal response of suspended nanowires to laser-induced thermal waves
}

\begin{abstract}
Thermal and photothermal effects play an increasing role at the nanoscale due to the general decrease of thermal conductances  and to the increasing role of interfaces. Here we present a non-contact optomechanical analysis of the thermal and photothermal properties of suspended nanowires based on pump-probe response measurements: a probe laser measures the nanowire deformations and property changes caused by an intensity-modulated pump laser launching thermal waves propagating along the 1D conductor. The analysis of the dominant photothermal contributions to the nanowires response in the spectral and spatial domains allows in particular to quantify the interfacial contact resistance, to detect its internal optical resonances  and to  image absorption inhomogeneities. Additionally, by exploiting the temperature-induced optical reflectivity changes of the nanowire, we  directly image the spatial structure of the thermal waves propagating within the nanowire.

Finally we investigate how those thermal waves are responsible for a dynamical modulation of the nanowire vibration frequency, in the resolved sideband regime,  providing  novel analytical tools to further inspect the structural properties of nano-optomechanical systems with a large signal to noise ratio.  
Those methods are generic and critical to correctly understand  the photothermal dynamical back action processes and improve the intrinsic sensitivity of those ultrasensitive force probes.

\end{abstract}

\maketitle

\begin{figure*}[t!]
\begin{center}
\includegraphics[width=0.9\linewidth]{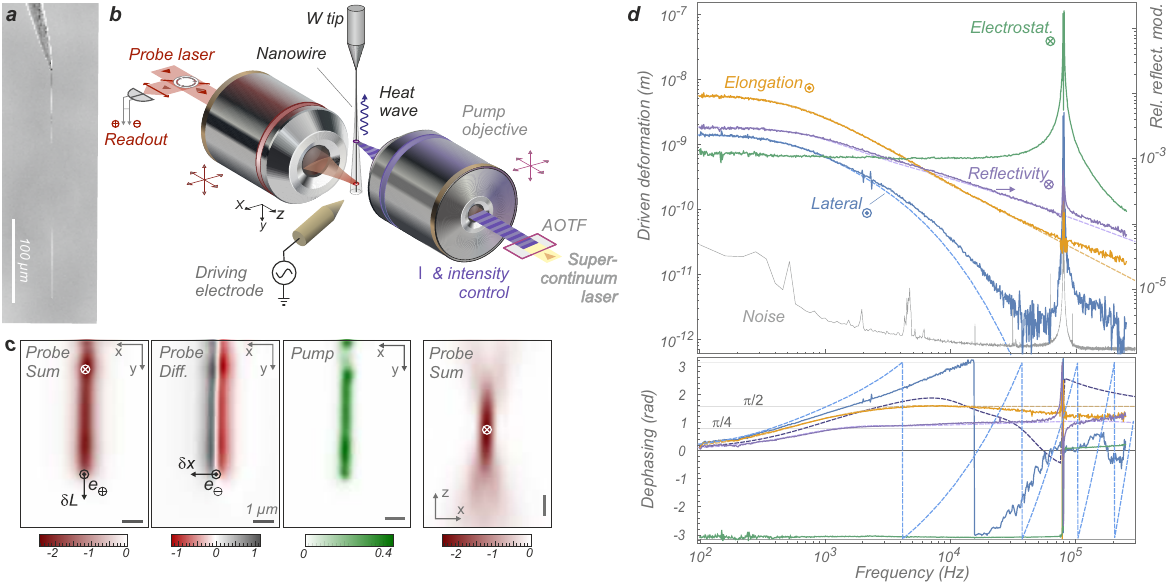}
\caption{
\textbf{Principle of the experiment.} {\bf a} SEM image of a suspended silicon carbide nanowire attached to a tungsten tip. {\bf b} Principle of the experiment: the nanowire vibrations are readout by a probe laser, whose reflection is collected on a split photodetector followed by a low noise amplifier providing the sum and difference of each photocurrents. A supercontinuum laser spectrally filtered and intensity modulated by an AOTF, serves as a pump laser to drive the nanowire via optical and photothermal actuation. The thermal waves induced by the time-modulated laser absorption are propagating along the nanowire up to the support tip. Moving the pump and probe laser positions allows  understanding and identifying the dominant driving mechanisms. {\bf c} Reflection images  obtained by scanning the probe (red) and pump (green) lasers in the vertical and horizontal planes with a $55\,\rm \mu m$ long nanowire, 220\,nm in diameter: when the laser waist is positioned at the nanowire extremity (marked by $\odot$), the difference/sum channels measure the lateral/longitudinal deformations of the nanowire, while the reflectivity changes can be measured on the sum channel  when the nanowire sits exactly at the optical waist (marked by $\otimes$). {\bf d} Typical response measurements in amplitude and dephasing obtained at those different measurement positions upon optical or electrostatic actuation. The nanowire elongation response  $\delta L [\Omega]$ can be modeled by a quasi perfect first order low pass filter (dashed lines). The lateral deformations $\delta x [\Omega]$ are dominated at low  frequency by the bilayer effect induced by the thermal wave arriving at the clamp area, caused by the different thermal expansion coefficients of the nanowire and its metallic support.  On the contrary, the laser reflectivity modulation $\delta R_{\rm opt}[\Omega]$ is caused by the amplitude of the thermal wave at the probe laser position and thus represents a local measurement channel.
}
\label{Fig1}
\end{center}
\end{figure*}

Enabling the conversion of a force into a measurable displacement, nanomechanical oscillators play a central role in force sensing at the nanoscale, providing extreme sensitivities and a great versatility of use. Force measurements require reading out their vibrations using  optical \cite{Fabre1994,Aspelmeyer2014,Mercier2017,Rossi2016,Ramos2012,Montinaro2014,Tavernarakis2018,Fogliano2021a} or electron-based  \cite{LaHaye2004,Lassagne2009,Moser2014,Nigues2015,Pairis2019,Khivrich2019} techniques, whose impact on the nanomechanical force probe needs to be carefully addressed in order to fully benefit from their exceptional sensing properties \cite{Pinard2008,DeLiberato2011}. In optomechanical systems, photon absorption represents an irreversible source of probe inefficiency, 
and is often found to be a limiting factor in those experiments: part of the probe signal can be absorbed in the resonator, causing a temperature increase and a heat wave that subsequently travels within the force sensor. 
The reduction in size of the oscillators enhances their mechanical response to detect ever smaller forces, but it also leads to an acceleration of those photothermal effects whose characteristic propagation time can become comparable to the vibration frequencies of the nanomechanical probes, all the more so at low temperatures, when the heat conductances vanishes. In this situation, the fluctuations of the probe field directly impact the resonator noise temperature, in addition to the more standard temperature increase due to the mean power absorbed. Combined to the somehow lower control of their material properties and to the increasing sensitivity to the interface imperfections, the relative impact of those photothermal effects dramatically increases at the nanoscale. The investigation of heat transport phenomena at this scale is a subject of research in itself, motivated in particular by the emergence of new regimes of heat propagation \cite{Seol2010,Pop2010,Swinkels2015,Tavakoli2017,Tavakoli2018} where the mean free path of the heat carriers compares with the system dimensions.
Among others, the deviation from Fourier laws, the transition from diffusive to ballistic regimes of heat propagation, the appearance of non-standard radiative heat transfer at the nanoscale \cite{Rousseau2009,Greffet2017}, the increasing  role of interfaces and even the possibility to define a local temperature are fundamental questions that arise in this field \cite{Ziman2001,Tavakoli2018,Vakulov2020,Chen2021,Zardo2019,Ziman2001},  which are directly relevant to the operation of those ultrasensitive electro-optomechanical force probes \cite{Zhou2019,Fogliano2021a,Briant2022,Cattiaux2021,Collin2022,Collin2023}. 

Despite being an often unwanted source of measurement backaction, photothermal effects are the core mechanism on which are built several nano-optical and nano-optomechanical experiments and sensors \cite{Allegrini1992,West2023,Kanellopulos2025}, with applications in single particle detection and exploration of their interactions to the light field \cite{Crut2014,Vella2018,Chen2022},  bolometers \cite{Blaikie2019} and infrared detection. Photothermal effects are responsible for light-induced thermal deformations of the resonators, optical non-linearities such as bistabilities of the optical cavities used to probe their vibrations, or  dynamical optomechanical backaction \cite{Metzger2004,DeLiberato2011}. They also impact the oscillators' frequency stability and thus practically limit their sensing capacities.\\ 
This great variety of actuation and perturbation mechanisms often render their distinction difficult to handle, particularly at the nanoscale. 
The miniaturization of the mechanical resonators increases their sensitivity to imperfections at their surface and the role of their interfaces to the external reservoirs, modelled as contact resistances through which the heat deposited by the probe flows.
Such interfacial effects are difficult to evaluate without spatial resolution and are often casted into an imperfect (lower) thermal conductivity of the devices under test.\\

It is worth mentioning that a  great variety of temperature-informative sensing techniques has already  been developed and implemented on micro and nanoresonators, such as Raman signal thermometry  \cite{Braun2022, Doerk2010, Lee2006, Abel2007,Hsu2008,Deshpande2009}  that exploits either the Stokes/anti-Stokes sideband asymmetries or their spectral shifts with the temperature changes, scanning thermal probe microscopy \cite{Majumdar1999,Gomes2015}, or direct thermal noise thermometry \cite{Mamin2001,BleszynskiJayich2008,Tao2016,Heritier2018,Cattiaux2021,Shaniv2023}. The dynamical investigation of the  photothermal effects through pump-probe methods has been largely used in optomechanical settings, and served to investigate the photothermal response of macroscopic cantilever \cite{Allegrini1992} and mirror \cite{DeRosa2006}, of graphene  and other 2D material membranes \cite{Schwarz2016,Schwarz2016a,Dolleman2018,Blaikie2019,Dolleman2020}, of high-Q microcavities \cite{Primo2021} or the impact of a superfluid He layer \cite{Riviere2013} and are very informative on the physics at play inside the resonator, while photothermal reflectivity changes were also exploited to detect sub-wavelength sized nanoparticles \cite{Crut2014,Vella2018,Chen2022}.

In this work, we  investigate the photothermal properties of a simple nanomechanical resonator, a suspended silicon carbide nanowire. A pump laser beam is focused on the nanowire and serves as a time-modulated local heat source,  which is launching thermal waves propagating within the nanowire. Those thermal waves generate minute deformations the resonator, change its optical properties and its mechanical stiffness, are responsible for photothermal forces and also modify and modulate its mechanical resonance frequencies. The lateral and longitudinal vibrations  and deformations of this quasi mono-dimensional nano-pendulums are readout using  an independent focused probe laser beam, in a configuration similar to the one employed to probe the lateral forces experienced by those ultrasensitive force sensors \cite{Gloppe2014,Mercier2017,Mercier2018,Fogliano2021a}. Moving the pump laser at different positions along the nanowire, we realize spatially resolved pump-probe response measurements to provide a quantitative analysis of its photothermal response.\\
We demonstrate that it is possible to achieve a rather complete and quantitative understanding and modeling of the dominant photothermal phenomena, which allows  reporting  on a few new observations and measurement techniques developed for this purpose. They are based on several key ingredients. First, with the possibility to displace the heat source within the resonator, we gain access to the spatial and temporal profiles of the thermal waves. In addition, continuously tuning the pump laser wavelength gives access to the spectral dependence of the light-nanowire interaction, demonstrating the contribution of the internal Mie resonances to the absorption efficiency, while also revealing cross-section self modulation forces.\\
Finally, we developed a novel methodology to investigate the dynamical frequency shifts of the nanowire eigenmodes induced by the time-modulated pump laser. They originate from the temperature dependence of the nanowire stiffness and from the deformations of the resonator caused by the thermal waves, but also from the force field gradients produced by the pump laser. The capacity to investigate those dynamical frequency shifts in the spatial and temporal domains allows to further understand the internal dynamics of the nanowire vibrations and how the balance between deformation and strain fields, as well as external force fields gradients set the frequency of a nanomechanical device.
The proper understanding of the internal dynamics of the thermal waves, and more generally of the dynamical optomechanical response of the nanowires is a prerequisite to correctly understand the backaction process and in particular how the probe and environment fluctuations impact the oscillator fluctuations.\\

\begin{flushleft}
{\bf Setup and  measurement principles \\}
\end{flushleft}

This study was conducted on two different experiments, operating at room  and cryogenic \cite{Fogliano2021a} temperatures, but we  mainly report here on the results obtained in the room temperature experiment which provides a better degree of control and a somehow simpler phenomenology for what concerns the heat propagation within the nanowires, which remains purely diffusive. As sketched in Fig.\,1b, the focal spots of a probe laser and a pump laser can be independently positioned along the length of a suspended nanowire. The reflection of a probe laser is employed to record the nanowire vibrations and mechanical deformations \cite{Mercier2017} induced by the pump laser whose intensity can be time-modulated in order to analyze to the temporal and spectral dependencies of the nanowire response. A voltage-biased metallic electrode provides a spectrally flat reference electrostatic force  \cite{Mercier2017} to actuate the nanowire and track its frequency shifts.

The nanomechanical resonators employed are suspended silicon carbide nanowires \cite{Gloppe2014,Mercier2017,Mercier2018,Fogliano2021,Fogliano2021a} attached to a sharp tungsten tip using an annealed carbon glue, as shown in Fig.1a. They have typical diameters $2R$ in the 100-400\,nm range and lengths $L$ extending up to a few hundreds of $\rm \mu m$.
The nanowire fundamental eigenfrequencies $\Omega_{\rm m}/2\pi$ are oscillating are in the kHz-MHz range depending on their geometry, with quality factors around 1000-10\,000 in vacuum at room temperature, increasing up to 100\,000 at cryogenic temperatures \cite{Fogliano2021a}. Their relatively large refractive index (2.7) allows them to sustain internal optical resonances in the visible light domain, referred to cylindrical Mie resonances \cite{Bohren1983}, which  spectrally structure their interaction with the light fields, as we explore bellow. \\
The probe laser can be positioned at different positions along the nanowire. In force sensing experiments we probe the lateral vibrations of the nanowire and we position the laser a bit above its vibrating extremity (marked by $\otimes$ in the reflectivity maps of Fig.\,1c). There the lateral deformations of the nanowire create an unbalance between the fluxes collected on each quadrant of the split photodetector placed in reflection. In this work, we will also position the probe laser waist slightly below the nanowire (marked by $\odot$ in Fig.\,1c), where its temperature-induced elongation causes its extremity to penetrate more into the probe laser, thus increasing the amount of reflected light collected on the sum channel of the dual photodetector. The spatial sensitivities of those measurement channels are experimentally determined by iteratively moving the nanowire using a calibrated piezo-stage by known quantities in all xyz directions. They reach a few MV/m for the probe powers typically employed (around $10-100\,\rm \mu W$ at room temperature), which allows  detecting its vibrations and deformations with a large signal to noise (see Fig.\,1d), not only at the mechanical resonance frequencies, but also at lower frequencies where the dynamics of the thermal waves will be analyzed.\\
Additional information on the internal temperature of the  nanowire can be extracted from its vibration noise temperature \cite{Fogliano2021a}, but this requires a rather large optical heating (typically larger than 10\,K at room temperature) to carry meaningful information. Alternatively, as presented in the last section, the nanowire vibration eigenfrequencies are very sensitive to temperature changes, mainly because of the material softening observed at higher temperatures  \cite{Yibibulla2021,Boyd2013,Pottier2021},  and we  explore how the light-induced dynamical frequency shifts can be exploited to obtain novel information on the interaction of internal thermal waves with the nanowire dynamics, complementary to the deformation measurements.\\
The pump laser can also be independently positioned along the nanowire using a second XYZ piezo stage while its intensity is controlled and time-modulated using an acousto-optical modulator. It is produced either by a 532\,nm doubled YAG laser or by a Leukos supercontinuum laser source followed by an acousto optical tunable filter (AOTF) from A-A optoelectronics. The latter allows tuning the pump wavelength from 450 to 750 nm, with a typical spectral linewidth of 0.5-1\,nm, thus crossing the optical gap of silicon carbide (around 515\, nm for the 3C allotropic phase). Modulating  the pump intensity at a frequency $\Omega/2\pi$ according to $P(t) =  P_0 +\delta P_0 \cos\Omega t$, generates a modulated optical force which can drive the nanowire into motion, but also a modulation of the absorbed light, generating heat waves propagating along the nanowire.\\ 
The probe laser will then serve to probe the optomechanical and the photothermal responses of the nanowire using locked-in demodulators. In the following, we record and analyze the different dynamical responses of the nanowire, and in particular the spectral dependence of the driven nanowire elongation $\delta L[\Omega]$, lateral deformations $\delta x[\Omega]$, of its reflectivity changes $\delta R_{\rm opt}[\Omega]$  and of the dynamical frequency shifts $\delta \Omega_{\rm m}[\Omega]$ caused by the optical and temperature modulation. 
Changing the modulation frequency then allows adjusting the wavelength and propagation length of the thermal waves, which are critically damped within the nanowire in the diffusive regime. As such, the capacity to move the pump laser position along the nanowire and vary the modulation frequency provides spatial and temporal degrees of freedom to analyze the propagation of the photothermal waves. This helps to better understand the origins of the different photothermal mechanisms at play within the nanowire and obtain spatial information on its internal structure.\\

Typical response measurements are shown in Fig.1d, where we report the amplitude and the dephasing of the complex demodulated signals, defined according to $\delta x(t)= {\rm Re}(\delta x[\Omega] e^{-i\Omega t})$ where the dephasing is measured relatively to the  excitation signal (pump laser optical intensity or electrostatic modulation) (a $+\pi/2$ dephasing corresponding to a response delayed by a quarter of the modulation period).
On the lateral deformation measurements $\delta x[\Omega]$ (blue curves), one can first identify the characteristic peaked response of the two  quasi degenerated fundamental eigenmodes oscillating around 80\,kHz along  two perpendicular transverse directions.
The mechanical spectral response of the nanowire  $\chi[\Omega]= {1}/{M_{\rm eff} (\Omega_{\rm m}^2-\Omega^2-i \Omega \Gamma_{\rm m})}$ for a degenerated nanowire is quasi-flat at frequencies smaller than the first resonances, as verified under electrostatic actuation (green curves), which creates a frequency-independent reference force.\\
The nanowire can be efficiently driven by optical forces \cite{Gloppe2014}, especially when the pump laser is positioned close to its vibrating extremity where it is maximally  sensitive to a localized force actuation, but it can also be excited by photothermal mechanisms, the more prominent lateral contribution being in general caused by the thermal deformations of the clamping area:  the incoming thermal wave changes the contact temperature,  and the differential thermal expansion coefficients of the nanowire and its support generates a lateral actuation of the nanowire.\\  
As such, the response cutoff observed at lower frequencies (around 500\, Hz here) hints at a spectral change in the photothermal actuation mechanism, connected to the characteristic diffusion time of the heat waves. At low modulation frequencies, the diffusive thermal waves propagate along the entire nanowire, while they become more and more localised around the pump laser source  when the modulation frequency increases, thus decreasing the photothermal actuation strength at the clamping.  The lateral photothermal deformations $\delta x[\Omega]$ thus present a peculiar response profile characteristic of a delayed diffusive mechanism ( $e^{-\sqrt{i\Omega \tau}}$ see below)  associated to the heat diffusion time from the pump to the clamping area ($\tau \sim L^2/D$).\\
On the contrary,  the elongation measurements $\delta L[\Omega]$ (orange curves) behave as a quasi-perfect first-order low-pass filter on a few horizontal and vertical decades.\\ 
Finally, at the measurement position $\otimes$, the sum channel is not sensitive at first order to lateral deformations, but only to laser-induced reflectivity changes $\delta R_{\rm opt} [\Omega]$. The pump laser modulates the  temperature of the nanowire and thus its refractive index causing a modulation of the probe laser reflectivity. This provides a measurement channel of the local light-induced temperature modulation, featuring a characteristic $1/\sqrt{\Omega}$ slower slope and a dephasing of  $\pi/4$ at modulation frequencies larger than the thermal cutoff when both lasers are superimposed (purple curve).
In the room temperature experiments, the temperature modulation involved typically range from 0.01 to 10 K, while still preserving a large signal-to-noise above the detection background (grey curve in Fig.\,1d, 1\,Hz resolution bandwidth). 
We now present a modeling of the heat wave propagation within the nanowires to account for those generic observations.\\

\begin{flushleft}
{\bf Modeling \\}
\end{flushleft}
At room temperature the heat propagation is mainly ensured by phonons propagating in the system. The mean path length of the ones oscillating at the dominant black body frequency falls in the nanometer range \cite{Tavakoli2018} and is thus far smaller than the nanoresonator's smallest dimensions.  A local temperature  $T(\vv{r},t)$ can  thus be defined at any position $\vv{r}$ within the nanowire and the heat propagation can be described by a diffusion equation, involving the material conductivity, density and heat capacity,  $\kappa$, $\rho$ and $C$ respectively. The heat waves diffusively expand with a heat diffusion coefficient  of $D=\kappa/\rho C\approx 10^{-5}\,\rm m^2/s$ and the characteristic time taken to spread across the nanowire thickness  ($R^2/D$) falls in the ns range, which is way faster than all the other thermal and mechanical time scales involved in our experiments. The temperature evolution along the nanowire $T(y,t)$ thus follows a 1D heat propagation equation given by:
\begin{equation}
\rho C \partial_t T + \partial_y( - \kappa \partial_y T ) = A_{\rm abs} v_0(y) \left ( P_0 + \delta P_0 \cos \Omega t \right).
\end{equation}
Here we neglected the lateral losses since we operate in vacuum ($< 10^{-3}\,\rm mbar$) and radiative emission remains very small compared to the heat flux traveling within the nanowire (see SI). $A_{\rm abs}$ is the local light absorption coefficient, which can vary with the pump position due to spatial inhomogeneities and with the pump wavelength, as we will see below. The pumping term in the right hand side of the equation depends on the position of the laser and on the internal optical profile within the nanowire. So as to maintain a simple analytical formulation,  in most of the following discussions,  we will employ a Dirac-like optical heating profile:
$v_0(y)= \ \delta (y-y_0)/\pi R^2$, where $y_0$ is the pump laser position. The stationary solution of this equation can be expressed as
$T(y,t) = T_{\rm stat} (y)+ {\rm Re} \left(\delta T[y,\Omega] e^{-i\Omega t}\right)$,
where the first term represents the static temperature profile, induced by the mean absorbed optical power $A_{\rm abs} P_0$ and $\delta T[y,\Omega]$ is the complex representation of the driven temperature response generated by the time modulated pump intensity ${\rm Re}\left(  \delta P_0\, e^{-i\Omega t}\right)$. The dephasing of the temperature modulation is thus computed in reference to the optical modulation.
In a first step, we will use the following boundary conditions:  $\partial_y T( L, t)=0$ and $T(0,t)= T_0$, which respectively describe the absence of heat loss at the nanowire vibrating extremity as well as a perfect thermal contact at the clamping point ($y=0$), $T_0$ being the experiment temperature. Under those assumptions, the static temperature profile can be expressed as
$T_{\rm stat} (y) = T_0+ \frac{P_{\rm abs}}{\pi R^2 \kappa} \left( y+ \Theta(y-y_0) (y_0-y) \right) $, $\Theta$ being the Heaviside function. It linearly increases from the clamping to the pump laser position and remains constant thereafter (see Fig\,4a). When the pump laser is positioned at the extremity of the nanowire, the maximal temperature increase is of $T_{\rm stat} (L)-T_0 = P_{\rm abs}  R_{\rm nw}$, where we have introduced the nanowire thermal resistance $R_{\rm nw}\equiv L/ {\pi R^2 \kappa}$, which amounts to approx. $3\times 10^7\,\rm K/W$ at room temperature for a $100\,\rm\mu m$ long and 100\,nm radius nanowire. The dynamical temperature profile is given by:

\begin{equation}
\frac{\delta T [y,\Omega]}{ \delta P_{\rm abs}} =\frac{R_{\rm nw}}{2 k L} \left(
\begin{array}{l}
 \frac{e^{k  (y_0-  L)}+ e^{k ( L- y_0)}}{e^{k L}+ e^{-k L}}
\left(e^{k y}-e^{-k y}\right)\\
+
\Theta(y-y_0) \left( e^{k ( y_0- y)}- e^{k ( y- y_0)}\right)\\
\end{array}
\right),
\end{equation}
where  $\delta P_{\rm abs}= A_{\rm abs} \delta P_0$ and $k \equiv \sqrt{-i \Omega/ D}$ represents the characteristic critically damped wave-vector of the thermal waves diffusing within the nanowire. Varying the modulation frequency will thus be a mean to explore different characteristic sizes within the nanowire. For modulation frequencies smaller than $\Omega_{\rm th}\equiv D/L^2 $, the thermal wave expands all along the nanowire with a spatial profile similar to the static profile given above, while at larger modulation frequencies the thermal waves tend to stay localized under the heating laser spot ( see Fig.\,4a). Below we will directly image those thermal waves through the reflectivity measurements and verify the good agreement with those expressions.\\

The photo-thermally-induced elongation of the nanowire is given by $\delta L[\Omega] = \int_0^L \alpha_L\  \delta T[y,\Omega] dy$, where $\alpha_L\approx 3\times 10^{-6}\,\rm K^{-1}$ is the material thermal expansion coefficient.  Its analytical expression is given in the SI. When the pump laser is positioned close to the nanowire vibrating extremity ($y_0\rightarrow L$), the  dynamical elongation can be  well approximated at low frequencies by
\begin{equation}
\frac{\delta L[\Omega]} {L }
\approx \alpha_L \frac{R_{\rm nw}}{2}
\frac{\delta P_{\rm abs}}{1-i \Omega L^2/2D},
\label{eq.dL}
\end{equation}
which behaves as a first order low-pass filter with a cutoff pulsation of $2D/L^2$,  which is consistent with the measurements shown in Fig.\,1d (orange curves).\\

As visible in the SEM images of Fig.\,4a, the complexity of the tip-nanowire interface prevents from realizing a realistic modeling of its thermal response.
However it can be safely assumed that the photothermally-induced lateral deformations produced at the tip are proportional to the temperature modulation arriving at the clamping point. Up to here we have  assumed a perfect thermalization of the nanowire with the bath,  so we will evaluate it at a position $y_{\rm bl}$ slightly inside the nanowire,   whose exact location does not matter in the coming evaluation as long as it remains small compared to the shorter thermal wavelength experimentally generated ($k y_{\rm bl} \ll 1$). Below we will introduce a more refined model of the thermal properties of the clamping, but this will not change this evaluation criterion. The bilayer deformations can thus be expected to evolve proportionally to $\delta T[y_{\rm bl},\Omega] / \delta T[y_{\rm bl},0]={\cosh k (L-y_0) }/{\cosh kL}$, with a prefactor that depends on the specific arrangement of the clamping area and may vary from one nanowire to another, in strength and in orientation.  When the pump laser is  positioned close to the nanowire vibrating extremity, and when the modulation frequency is larger than the thermal cuttoff ($k L\gg 1$), this ratio scales as  $e^{-k L}= e^{-(1-i)\sqrt{\Omega/2DL^2}}$, its amplitude rapidly decays with the modulation frequency, while the dephasing  increases with the modulation frequency, reproducing the behaviour reported in Fig.\,1d. \\

Those two characteristic behaviors are in good agreement with the experimental observations of the elongation and lateral deformations reported in Fig.\,1d.
We will now turn to a more quantitative analysis of the measurements and first discuss the contact interface.

\begin{figure}[t!]
\begin{center}
\includegraphics[width=0.99\linewidth]{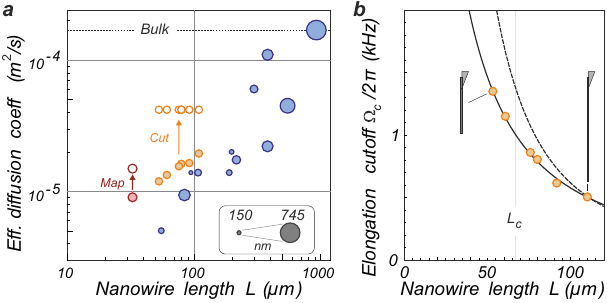}
\caption{
\textbf{Effective diffusion coefficients and contact thermal resistance} {\bf a} Effective diffusion coefficients derived from the measured elongation thermal cutoff as a function of the nanowire suspended length. The shorter nanowires present a significantly lower thermal cutoff that deviates from the bulk value indicated by a dashed line. The diameter of the nanowires is indicated in the inset. Open circles represent the diffusion coefficient compensated from the thermal contact resistance. The red points are derived from the analysis of the spatial dependence of the response measurements (see below)  while the orange points are derived from the experiment of panel {\bf b} which represents  the evolution of the thermal elongation cutoff measured on a single nanowire, while progressively reducing its vibrating length via laser cutting. The dashed line is the $1/L^2$ dependency expected in absence of contact resistance, while the full line represents $2D/L^2/(1+2 L_c/L)$ with $L_c = 66\,\rm \mu m$ and $D= 4.2\times 10^{-5}\,\rm m^2/s$.
}
\label{Fig2}
\end{center}
\end{figure}

\begin{flushleft}
{\bf Thermal contact resistance \\}
\end{flushleft}

Over the years, we accumulated response measurements realized on  multiple nanowires, with varying lengths and diameters. Each measurement of the thermal cutoff allows extracting an effective diffusion coefficient of the nanowire, which are reported in Fig.\,2a. We observe a generic trend that short nanowires present a smaller value compared to longer ones, for which it becomes comparable to the bulk value ($1.6\times 10^{-4} \,\rm m^2/s$ \cite{Levinshtein2001}). Since the diameters of the different nanowires tested were not correlated to their length, it was difficult to believe that this trend was connected to some internal properties of the nanowires.\\
We  then realized an experiment where we progressively laser cut the same nanowire from 110 to $53\,\rm \mu m $ (using optical powers beyond 10\,mW in vacuum) and recorded photothermal response measurements after each cutting step (see Fig.\,2b). We observed that the thermal cutoff frequency does not follow the expected $D/ L^2$  behavior, which pointed towards the existence of a non-perfect thermal contact.
This imperfectness of the thermal contact can be modeled by a thermal resistance, $R_c $ (in K/W),  positioned between the heat bath and the nanowire, as sketched in Fig.\,4a.
If we neglect the thermal inertia of the contact area, the temperature and the heat flux at the nanowire clamping extremity  are connected through $T(0,t)-T_0 = R_c \pi R^2  \kappa \left. \partial_y T \right|_0 $. When the pump laser is positioned at the nanowire vibrating extremity, the static temperature gradient becomes $T_{\rm stat}(L)- T_0 =  (R_{\rm nw}+R_c) P_{\rm abs}$, which simply translates the fact that the contact resistance adds up to the nanowire resistance. The dynamical boundary condition becomes $\delta T[0,\Omega]= R_c \pi R^2 \kappa \,\partial_y\delta T[0,\Omega]$, which leads to the modified temperature  and elongation profiles:
\begin{widetext}
\begin{equation}
\frac{\delta T [y,\Omega]}{ \delta P_{\rm abs}}=
\frac{R_{\rm nw} }{2k L}\left(
\frac{e^{k (L-y_0)}+e^{k (y_0-L)}}
{e^{kL}(1+ k L_c)+e^{-k L} (1-k L_c)}
\left((1+k L_c) e^{k y}-(1-k L_c) e^{-k y}\right)
+\Theta(y-y_0) (
e^{k (y_0-y)}-e^{k (y-y_0)}
)
\right)
\label{eq.dT-Rc}
\end{equation}
\end{widetext}
\begin{widetext}
\begin{equation}
\frac{\delta L[\Omega]/L}{\delta P_{\rm abs}}=\frac{\alpha_L R_{\rm nw}}{2 k^2 L^2} \left(
\frac{e^{k  (y_0-  L)}+ e^{k ( L- y_0)}}
{e^{kL}(1+ k L_c)+e^{-k L} (1-k L_c)}
\left((1+kL_c)e^{k L}+(1-kL_c)e^{-k L}-2\right)
+
2- e^{k ( y_0- L)}- e^{k ( L- y_0)}
\right),
\label{eq.dL-Rc}
\end{equation}
\end{widetext}
where we introduced the contact length $L_c \equiv L R_c/R_{\rm nw}$, which converts the contact resistance into an equivalent  supplemental nanowire length (see Fig.\,4a).
When the pump laser is positioned at the extremity of the nanowire, the dynamical elongation response can be approximated by
\begin{equation}
\frac{\delta L[\Omega]}{L }
=
\frac{ \alpha_L R_{\rm nw} }{2 }(1 + 2 R_c/R_{\rm nw} )\frac{\delta P_{\rm abs}}
{1-i\Omega\frac{L^2}{2 D}(1 + 2 R_c/R_{\rm nw} )}.
\end{equation}
The contact resistance thus increases the static elongation since its presence leads to an increase of the temperature within the nanowire, but is also responsible for a reduction of the thermal cutoff
\begin{equation}
\Omega_c^{\rm th} = 2D/L^2/(1+2 L_c/L)
\label{eq.OmegacvsLc}
\end{equation}
by the same amount. As such, when cutting the nanowire, we will observe a deviation from the $1/L^2$ law, especially when the nanowire length becomes comparable to $L_c$. This expression allows  perfectly adjusting the measured evolution of the thermal frequency cutoffs as a function of the nanowire length and to determine the contact resistance (see Fig.\,2b). The equivalent nanowire contact length $L_c$ varies from one nanowire to another, but is in general of a few tens of microns (comparable to the length over which we glue the nanowires on their metallic support).  This  observation also explains why the longest nanowires are less affected by the imperfection of the clamping, leading to a larger effective diffusion coefficient, as reported above. Such a method to quantify $R_c$ is obviously very destructive for the nanowire, but we will show in the following that the analysis of the spectral and spatial dependencies of the nanowire photothermal response can provide an alternative, non-invasive, determination of the thermal contact resistance, while also providing refined structural information on the nanoresonator.

\begin{figure}[b!]
\begin{center}
\includegraphics[width=0.95\linewidth]{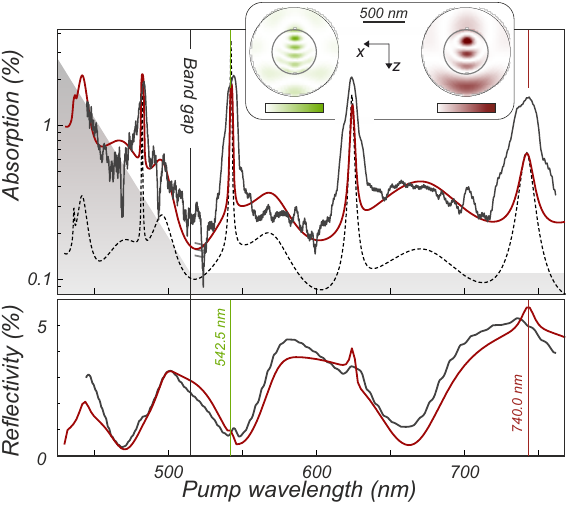}
\caption{
Spectral dependence of the absorption and reflection coefficients of the pump laser of a 555\,nm diameter nanowire. The former is extracted from a dynamical elongation measurement  (at 80\,Hz) realized with the red probe laser, while the latter is directly demodulated from a photodiode placed in reflection (60\,s sweep duration,  10\,Hz resolution bandwidth). The Mie resonances strongly structure the absorption spectrum of the nanowire.
The measurements are compared to the theoretical prediction (red line)  which accounts for the enhancement of the intra-nanowire light field caused by the Mie resonance (red dashed line, see SI), combined with a bulk absorption spectrum  whose spectral shape is shown as the darkened area, associating a flat background absorption (residual penetration depth of $ 40\,\rm cm^{-1}$) to a gap-like behavior below 515\,nm. We also account for the changes in the illumination conditions during the color sweep (see SI). The insets represent the normalized light intensity distribution around the nanowire for two different resonant pump wavelengths, the pump laser propagating along the z-axis.
}
\label{Fig3}
\end{center}
\end{figure}

\begin{flushleft}
{\bf Optical and spatial analyses of the photothermal response \\}
\end{flushleft}
The strength of the optically induced heat source depends on how the pump laser is absorbed by the nanowire. Silicon carbide is a  large bandgap semiconductor, falling around 525\,nm (2.36\,eV) for bulk material for the 3C allotropic phase we employ. To obtain structural information on the nanoresonator, we sweep the optical pump wavelength in the visible domain with the AOTF, filtering and modulating the intensity of the super-continuum laser. The pump laser wavelength is continuously adjusted from 450 to 750\,nm by sweeping the AOTF driving radio frequency tone, with a spectral linewidth of 0.5-1\,nm. The pump intensity is controlled and modulated by changing by the radio-frequency amplitude, with a modulation bandwidth around 500\,kHz, which is independently calibrated on a reference photodiode positioned before the vacuum chamber.\\
We first position the pump laser close to the vibrating extremity of the nanowire and measure its photothermal response at a fixed modulation frequency (around 80\,Hz here) while continuously sweeping the pump laser wavelength $\lambda_{\rm pump}$. The spectral evolution of the elongation signal of a relatively thick nanowire ($2R= 550\,\rm nm$ is shown in Fig. 3a. It presents multiple peaks, relatively sharp (down to 10\,nm linewidth) corresponding to the successive internal Mie resonances of the cylindrical nanowire, as well as a global increase observed at shorter wavelengths, falling beyond the SiC bandgap. The knowledge of the injected optical power and of the material expansion coefficient allows extracting the absorption coefficient $A_{\rm abs}$ and its wavelength dependence using equation (\ref{eq.dL}).  When the pump wavelength is resonant with the Mie resonances, the cylinder is optically resonant, and a larger amount of light enters the nanowire, thus enhancing the overall absorption efficiency. It can be computed using the analytical description of the light-nanowire interaction \cite{Bohren1983, Reigue2023} (see SI). Such an analysis also provides a precise determination of the local nanowire radius (within a few nanometer resolution) by adjusting the Mie resonance wavelengths. The pumping efficiency also depends one the light polarization since TE and TM modes are not resonant at the same wavelengths (not shown). Finally, by combining the Mie contributions with a gap-like absorption below 515\,nm and a flatter residual absorption above, we can qualitatively reproduce the measured absorption coefficient, see Fig.\,3a and SI. The theoretical peaks are systematically narrower than the measured resonances, especially at larger wavelengths. This could be a consequence of the nanowire surface rugosity (around 1-10 \,nm depending on the presence of an oxyde crust) which is more impacting the less confined lower order modes.\\

The map shown in Fig.\,4e, represents the elongation and lateral deformations measured at a fixed quasi-static modulation tone (approx 1\,kHz), when scanning the pump laser position along a shorter nanowire. This measurement highlights locations where the nanowire absorbs the pump light far more efficiently (a ten-fold increase can be observed) than in other areas. The absorption spectra at those "hot spots" often present a slight shift of the Mie resonances and are in general associated to a local variation of reflectivity. We believe that most of the hot spots are due to a local diameter change (often due to a modification of the allotropic phase of SiC during the growth), which increases the light injection efficiency and thus  the local absorption coefficient.\\
This non-homogeneous absorption efficiency causes the amplitude of the nanowire photothermal response to strongly vary with the pump laser position and provides qualitative information on the internal structure of the nanowires.\\

\begin{figure*}[t!]
\begin{center}
\includegraphics[width=0.98\linewidth]{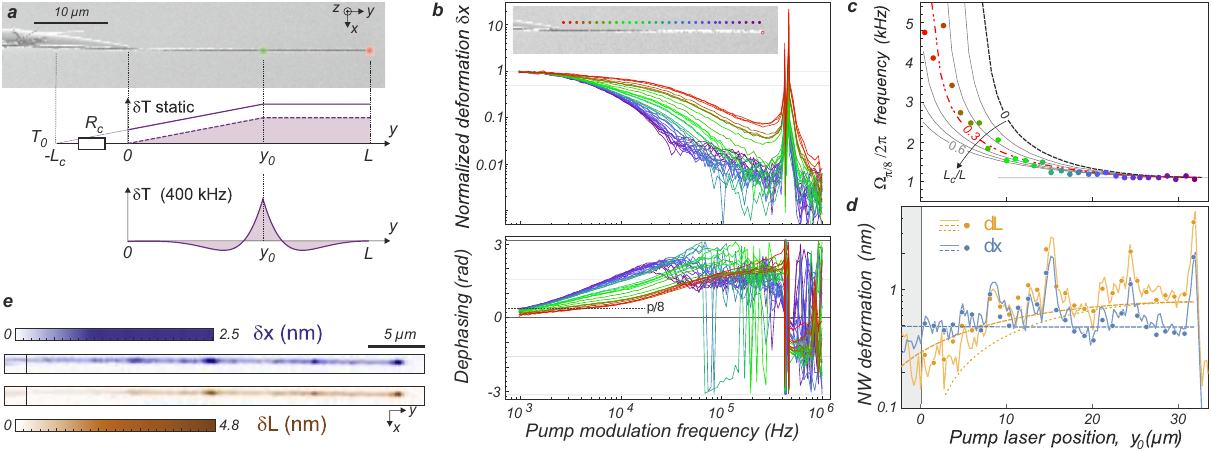}
\caption{
\textbf{Spatio-spectral maps and contact resistance} {\bf a} Sketch of the model superimposed on a SEM image of the $33\,\rm \mu m$ - 440\,kHz nanowire. The bottom plots represent the temperature profiles $Re (\delta T[y,\Omega])$ induced by the pump laser in the static and dynamical cases. The contact resistance, which can be modeled by a fictive additional nanowire length $L_c$ before the thermostat, modifies the temperature profile at low modulation frequencies,  when the thermal wavelength is comparable to the nanowire dimension.  {\bf b} Normalized lateral responses $\delta x[\Omega]/\delta x[0]$ measured for varying positions of the pump laser (shown as colored dots in the inset). The lateral deformations are dominated by the bilayer effect, which becomes faster (larger bandwidth) when the pump laser approaches the clamping area (red plots). {\bf c} Evolution of the cutoff frequency $\Omega_{\pi/8}/2\pi$ for which the lateral deformations are delayed by $\pi/8$ with respect to the pump power modulation, as a function of the pump position $y_0$. In absence of thermal contact resistance, the cutoff diverges towards the clamping position (black dashed curve). The measured evolution can be adjusted with numerical solutions of $\rm arg\left( \delta T[0,\Omega_{\pi/8}]/ \delta T[0,0]\right) = \pi/8 $ derived from eq.\,(\ref{eq.dT-Rc}), obtained for varying $L_c/L$  and $D$ coefficients.
The best fitting curve (dashed red) is obtained for $D=1.5\times 10^{-5}\,\rm m^2/s$ and $L_c/L=0.3$. The other curves are obtained by maintaining the effective diffusion coefficient  $D/(1+2 L_c/L)$ fixed to reproduce the 1100\, Hz cutoff measured at the vibrating extremity.
{\bf d} Comparison of the quasi-static elongation and deformation measurements realized for varying positions of the  pump laser. The full lines are extracted from the maps shown in panel {\bf e}, obtained by scanning the pump laser in the vertical xy-plane while using a fixed 1015\,Hz modulation tone. They  are superimposed with the results of the local response measurements realized at each discrete positions (shown in panel b and in the SI). The absorption inhomogeneities are responsible for the local variability observed. The generic trends show that when the pump laser approaches the clamping point, the elongation signal vanishes, while the bilayer mechanism remains almost constant. The dashed lines represent the theoretical expressions, using the parameters derived from the response cutoff analysis.
}
\label{Fig4}
\end{center}
\end{figure*}

\begin{flushleft}
{\bf Spatial mapping of the photothermal response}
\end{flushleft}

We now turn to a more quantitative analysis of the thermal waves propagating along the nanowire by realizing a complete set of response measurements at different pump laser positions along a $33\,\mu m$-long nanowire, shorter than the $50\,\rm \mu m $ piezo scanning range. Successive response measurements of the elongation and lateral deformation of the nanowire were realized by moving the pump laser position $y_0$ in $1\,\rm \mu m$ steps. The normalized driven lateral deformation, $\delta x[\Omega]/\delta x[0]$ are shown in Fig.\,4b, while the driven elongation measurements $\delta L [\Omega]$ are shown in the SI.
The elongation measurements present a low frequency response that is relatively invariant with the pump laser position, featuring the same first-order low-pass filter as explained above. The response amplitude varies with the local absorption changes and gets globally reduced when the pump laser approaches the clamping area.\\ 
On the other hand, the lateral deformations present a spectral structure that strongly varies with the pump laser position : when the pump laser is positioned at the vibrating extremity, the thermal waves have to travel a longer way across the nanowire before reaching the clamping area, where the lateral deformations are generated by the bilayer effect. This causes a rapid decrease in efficiency when the modulation frequency increases, as well as a rapid variation of the dephasing. 
On the contrary, when the pump laser is positioned close to the clamping position, the evolution of the dephasing with the modulation frequency gets flatter, as well as the response amplitude: the photothermal actuation is faster and more effective since the distance to travel is reduced. This acceleration is visible for all measurement positions, but somehow gets reduced when the pump laser approaches the clamping point: the last measurements are quasi identical, and the phase does not further flatten as one would  expecte when approaching the clamping position. This is a direct consequence of the thermal contact resistance, which effectively dissociates the clamping point from the actual reservoir position.\\
To put this observation in  more quantitative terms, we  report in Fig.\,4c the frequency at which the lateral response presents a $\pi/8$ dephasing compared to the static case (this value allows not to be affected by other driving mechanisms, such as optical forces, which are more visible at higher frequencies). We then compare our observations to the model predictions in presence and absence of thermal contact resistance using expression (\ref{eq.dT-Rc}).

In absence of contact resistance, $\Omega_{\pi/8}$  diverges when one approaches the clamping position ($y_0\rightarrow 0$)  which does not reproduce our observations, as pointed above. On the contrary, we obtain a good agreement with our measurements using a $L_c = 0.3 L$  and $D = 1.5\times 10^{-5}\,\rm m^2/s$.\\ 
A separate confirmation is also obtained in the analysis of the amplitude of the quasi-static elongation measured for varying pump laser positions, see Fig.\,4d, where we observe that the elongation does not drop down to zero when the pump laser reaches the clamping point (dashed orange curve in Fig.\,4d), but to a finite value (full orange curve). These results will also be further confirmed in the analysis of the dynamical frequency shifts realized on the same nanowire (see below).\\
As such, we have shown how it is also possible to extract the thermal contact resistance by analyzing the evolution of the response measurements with the position of the heat source in the nanoresonator.  Such a knowledge is essential to correctly understand and model the overall photothermal response of the system and can also serve to improve our understanding of the  thermal properties of the clamping area, which play an essential and increasing role when operating those ultrasensitive force sensors at dilution temperatures.\\

\begin{figure}[t!]
\begin{center}
\includegraphics[width=0.99\linewidth]{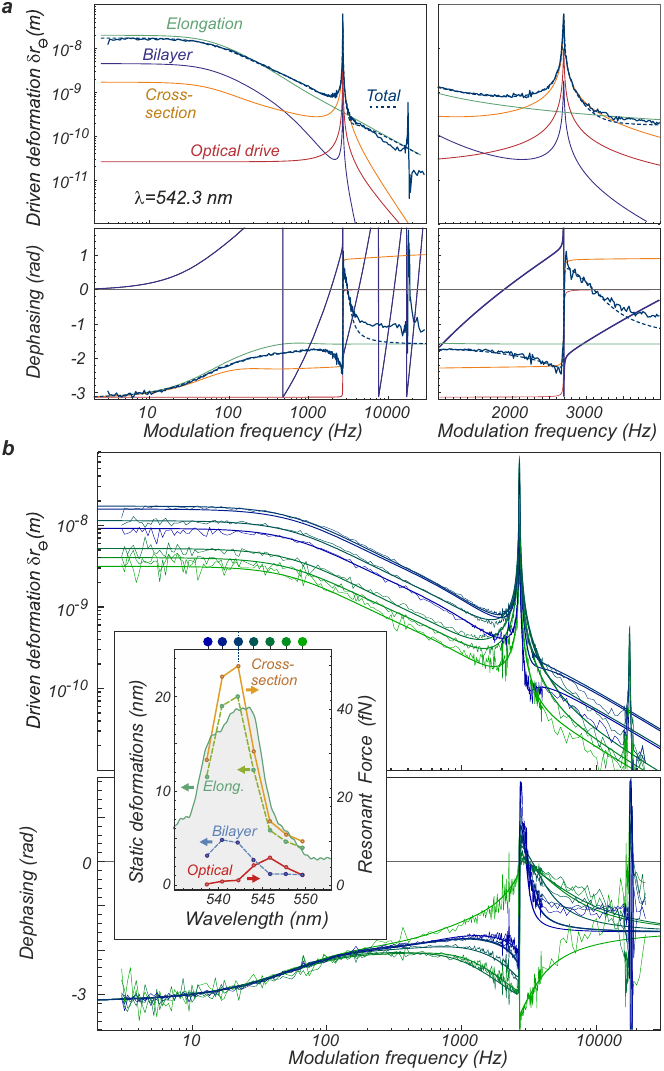}
\caption{
\textbf{Color dependence of the photothermal response measured across a Mie resonance.}  Lateral responses measured for different pump laser wavelengths, scanned across a higher order Mie resonance of the 550\,nm-nanowire (from 538.7 to 549.6\,nm). Each response can be well adjusted using  4 contributions that coherently add up (see text): optical force, bilayer and elongation processes, as well as laser induced cross section modulation, whose individual spectral contributions are shown in the upper plot (measured at 542.3 nm), the total response being the dashed blue line,  and whose fitting parameters are reported in the inset (see text). The elongation measured in Fig.\,3 while continuously scanning the pump laser wavelength is shown as a filled line for comparison. We have shown the static elongation and bilayer contributions (green and blue dashed lines), as well as the resonant contributions (at $\Omega_{\rm m}$) of the optical and cross-section modulation forces (red and orange dashed lines).
}
\label{Fig5}
\end{center}
\end{figure}

\begin{flushleft}
{\bf Complete fits - cross section modulation force\\}
\end{flushleft}
We now aim at adjusting the complete optomechanical response measurements to accurately account for and evaluate the different photothermal mechanisms at play within the system. To obtain a satisfying fitting result, we  had to introduce a third photothermal actuation mechanisms, the photothermally-induced optical cross-section self modulation. If we omit the vectorial and tensorial character of the optomechanical interaction, the optomechanical force exerted by the pump laser can always be written under the form  $F =  \sigma P $, where the cross section $\sigma$ models the light-nanowire interaction and $P$ is the pump laser intensity. When the latter is intensity-modulated, this generates a total modulated force of $\frac{\delta F}{\delta P} =  \sigma + P \frac{d \sigma}{dP} $. The first term represents the traditional optomechanical force, which instantaneously follows the light modulation. The second term arises when the nanowire  optical cross section is modulated by the light field, which can originate from photo-refractive effects (in phase with the light modulation on mechanical time-scales), but also from a photothermal contribution due to the temperature dependence of the nanowire refractive index and thus of the optical cross section  (the local nanowire thermal expansion contributes typically 10 times less than the refractive index change), as illustrated in the probe laser reflectivity measurements shown in Fig.\,1d (purple curves). The cross-section modulation force is thus proportional to the temperature changes below the pump laser spot, $\delta T[y_0,\Omega]$. Its magnitude is expected to be enhanced by the Mie resonances, which spectrally structure the light-nanowire interaction, so we conducted lateral response measurements on the thick nanowire (555\, nm diameter) at different optical wavelengths around 540\,nm, across the sharp Mie resonance visible in Fig.\,3. The response measurements shown in Fig.\,5  largely vary with the laser wavelength,  in agreement with the elongation measurement shown above, especially around the first mechanical resonances.\\
Those lateral responses can be well adjusted in amplitude and in phase by combining the three main photothermal contributions: the bilayer, the elongation and the cross section effects along with an instantaneous optomechanical force,  using the complex expression
\begin{equation}
\begin{array}{cl}
\delta x[\Omega] &
=
\left(
\delta x_{\rm bl} \frac{\delta T [0,\Omega]}{\delta T [0,0]}
+
 \delta x_{\rm \sigma} \frac{\delta T [y_0,\Omega]}{\delta T [y_0,\Omega]}
\right) \frac{\chi[\Omega]}{\chi[0]}\\
& \\
&
+\delta F_{\rm opt}\, \chi[\Omega]+
\delta x_{\rm el} \frac{\delta L [\Omega]}{\delta L [0]} \\
\end{array}
\label{eq.dxfit}
\end{equation}
where we employ the dynamical temperature profiles of eq.\,(\ref{eq.dT-Rc}) using $L=550\,\rm \mu m$,  $ y_0= 540\,\rm \mu m $, a measurement position at $y_{\odot}=547\,\rm \mu m$ and  $D = 4.5\times 10^{-5}\,\rm m^2/s$.  All contributions, except the elongation, present a resonant mechanical response and thus depend on the nanowire mechanical susceptibility $\chi[\Omega]$. We could neglect here the contact resistance contribution for this very long nanowire. The spectral responses of each components of eq.\,(\ref{eq.dxfit}) are separately shown  in Fig.\,5a, as well as the complete coherent response (dashed black line). The only fitting parameters varying between each measurements are the static real quantities  $\delta x_{\rm bl}$, $\delta x_{\rm el}$, $\delta x_{\sigma}$, as well as the optical force $\delta F_{\rm opt}$. The complete fits are shown as full colored lines, and the individual fitting parameters are shown in the inset, superimposed to the spectral elongation measurement taken from Fig.\,3.  The elongation and bilayer mechanisms play a dominant role at low frequencies and follow the optical dependence of the elongation measurement measured independently. Both mechanisms are less efficient at higher frequencies due to the thermal cutoffs,  which allows revealing the cross section modulation contribution, since it presents a weaker dynamical attenuation ($\Omega^{-1/2}$ vs $\Omega^{-1}$) as seen in Fig.\,1d. In the inset, we compared the optical force $\delta F_{\rm opt}$ to the resonant cross-section modulation force $\delta F_\sigma[\Omega_{\rm m}]=\frac{ \delta x_{\rm \sigma}}{\chi[0]} \frac{\delta T [y_0,\Omega_{\rm m}]}{\delta T [y_0,0]}$. The latter can largely dominate over the optical force when pumping at the Mie resonance.  We note that the ratio of both contributions largely depend on the measurement position within the optical waist since the optical force field is strongly structured in the waist area \cite{Gloppe2014}, especially  when optical gradient forces appear, but this aspect will be treated in greater detail elsewhere.\\
One can thus obtain a precise adjustment of the response measurements by adjusting the relative strength of the different photothermal responses and including the contribution of the cross-section self photothermal modulation. As already observed in the elongation measurements, Mie resonances strongly enhance the overall photothermal response of the nanowires, which can thus be tuned by adjusting the pump wavelength or polarization or the nanowire diameter.\\

\begin{figure}[t!]
\centering
\includegraphics[width=.45\textwidth]{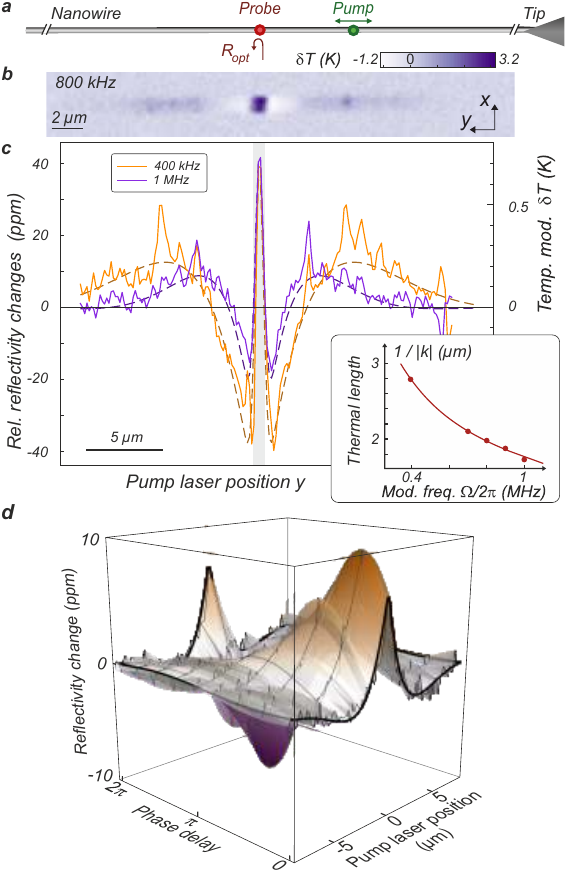}
    \caption{\textbf{Thermal wave imaging.} \textbf{a.} Measurement principle: we modulate the intensity of the pump laser (green) at frequency $\Omega/2\pi$ while scanning its position around the probe (red) laser's position $y_{\rm probe}$, located in the middle of the nanowire length. The response is taken on the sum channel, sensitive to reflectivity variations induced by the thermal wave. \textbf{b.} Map of the probe reflectivity changes converted in temperature change at $\Omega/2\pi=800\,$kHz: the critically damped thermal wave is centered  around the probe laser spot. \textbf{c.} Line cuts of the reflectivity changes taken along the nanowire's axis at two different modulation frequencies producing thermal waves featuring two different wavelengths. The dashed curves are the theoretical fits with the thermal model (see text) including an additional instantaneous contribution to the reflectivity changes - likely to be electrostriction - at the origin of the peak observed when both lasers superimpose. Inset: Evolution of the inverse thermal wavevector with the modulation frequency,  adjusted using  $1/|k|=\sqrt{D/\Omega}$ and $D= 1.9\times 10^{-5}\,\rm m^2/s$.
    \textbf{d.} Evolution in space and time of the driven photothermal wave (1\,MHz modulation), obtained by varying the display time $t$ (see text) over one period (NW29kHz, see SI). The surface with black contours is built from the fits, the  transparent one from the data.  
    }
    \label{Fig6}
\end{figure}

\begin{flushleft}
{\bf  Direct imaging of thermal waves}
\end{flushleft}

As shown above, the photothermal response measurements are affected by the imperfection of the thermal contact at the clamping, especially when using shorter nanowires.  The nanowire internal heat diffusion coefficient can be extracted from the proposed analyses, but it remains an indirect measurement.
Here we propose an alternative approach based on a direct imaging of the driven thermal wave profile within the nanowire, by making use of the thermally-induced reflectivity modulation mechanism  (illustrated in Fig.\,1d, purple curves).
At large modulation frequencies, the thermal wavelength is significantly shorter than the nanowire length ($|k | L\gg 1$) and when the laser is positioned in the middle part of the nanowire, the complex temperature profile of expression (\ref{eq.dT-Rc}) can be simplified to:
\begin{equation}
\frac{\delta T[y,\Omega]}{\delta P_{\rm abs}} = \frac{e^{- k | y- y_0|}}{2 \kappa \pi R^2\, k}  
\,\,\, {\rm with}\,\, k=\sqrt{-i\Omega /D}=e^{-i\pi/4} |k|,
\label{eq.thermalwave}
\end{equation}
which does not depend on the thermal contact properties nor on the nanowire length. This expression corresponds to the critically damped thermal wave, propagating away from the pump laser whose direct imaging thus provides a direct determination of the intrinsic nanowire heat diffusion coefficient $D$.\\
To do so, we positioned the probe laser at a fixed position in the middle region of the nanowire length, while launching high frequency  thermal waves with the pump laser from different positions along the nanowire. The chosen lateral position of the probe laser is similar to $\otimes$ in Fig.\,1c, where the sum channel is minimally sensitive to lateral deformation and serves to detect the reflectivity changes $\delta R_{\rm opt}$ induced by the propagating thermal waves diffusing within the nanowire. We chose modulation frequencies $\Omega/2\pi$ significantly larger than before, in the 400-1000\,kHz range to shrink the thermal wavelength below the piezo scanning range, while avoiding driving the higher order mechanical modes of the nanowire. 

The nanowire reflectivity changes are dominated by the local refractive index modulation $\delta n[\Omega]$ induced by the temperature modulation. The nanowire reflectivity strongly varies with the nanowire diameter and the optical wavelength due to the contribution of the Mie resonances \cite{Reigue2023}, but the expression  $\frac{\delta R_{\rm opt}[\Omega]}{R_{\rm opt}}\approx \frac{\delta n[\Omega]}{n}= \beta_n \delta T[\Omega]$, can however be qualitatively employed to estimate the corresponding temperature change, using $\beta_n = 5.7 \times 10^{-5}\,K^{-1}$, the tabulated thermorefractive coefficient of cubic SiC \cite{Powell2020}.\\

The results of those measurements are shown in Fig. 6bc: we scan the pump laser in the vertical XY plane around the probe laser position $y_{\rm probe}$ while modulating its intensity at a frequency $\Omega/2\pi$.  For each pump laser position, we record the amplitude and the dephasing of the driven reflectivity modulation (1\,Hz resolution bandwidth), of the order of a few tens of ppm,  and the thermal wave profiles shown in the plots correspond to  $ {\rm Re} \left(|\delta T[\Omega] | e^{i {\rm arg}\,  \delta T[\Omega]  -i\Omega t } \right)$ with $t=0$. This represents a snapshot of the reflectivity modulation at the moment when the intensity modulation is maximal and well reflects the critically damped nature of the photothermal wave. Varying the snapshot time $t$ over one modulation period ($\Omega/2\pi=1\,\rm \mu s$ here) then allows  visualising the spatio-temporal evolution of the driven thermal wave  (see Fig.\,4d). As in the "cave effect", the temperature wave is maximal under the pump laser spot for a time delay  of $(2\pi/\Omega)/8$ or a dephasing of $\pi/4$, as expected from eq. (\ref{eq.thermalwave}) associated to the photothermal dephasing of $\pi/4$ visible in Fig.\,1d purple curve.
The different lateral peaks visible in the signal amplitude of Fig.\,6c are due to absorption inhomogeneities along the nanowire, while the singularity visible in the center, where both lasers overlap has a different origin. It does not depend on the modulation frequency and the reflectivity changes are in phase with the intensity modulation,  meaning that the underlying mechanism is quasi-instantaneous at those modulation time-scales. It is very likely a signature of an electrostriction effect induced by the green pump laser \cite{Rakich2010}. This peak is not visible on all nanowires and depends on the presence of a Mie resonance at the pump/probe wavelengths  (see SI). \\
The reflectivity changes can thus be adjusted with the expression $\delta R_{\rm opt}/R_{\rm opt}= {\rm Re}\left( \beta_n \, \delta T[y_{\rm probe},y_0,\Omega] + \mu \, e^{-2(y_0-y_{\rm probe})^2/w_0^2}\right) $, the second term accounting for the local effects under the waist. The results obtained are equivalently well fitted using either the complete thermal profiles (using a Dirac (eq. (\ref{eq.dT-Rc})) or Gaussian heat source (see SI), including the thermal contact). If one seeks a model-free evaluation of the nanowire heat diffusion coefficient, it is possible to use the local approximation of eq. (\ref{eq.thermalwave}) and to directly extract the inverse thermal wavevectors  $1/|k|$  which are reported  in the inset of Fig.\,6c as a function of the modulation frequency $\Omega/2\pi$. The nanowire diffusion coefficient  can thus be directly extracted from a fit using the dispersion relation of the thermal waves: $|k|= \sqrt{\frac{\Omega}{D}}$. We obtained  $D = 1.9 \times 10^{-5}\,\rm m^2/s$,  a  value  slightly higher than the one obtained through elongation and deformation measurements ($D_{\rm eff}=1.6 \times 10^{-5}\,\rm m^2/s$), as expected from the above discussion on the thermal contact. From expression (\ref{eq.OmegacvsLc}), this corresponds to an equivalent thermal contact length of $L_c/L=\frac{D-D_{\rm eff}}{2 D_{\rm eff}} \approx 9\,\% $.\\

\begin{figure*}[t!]
\begin{center}
\includegraphics[width=0.99\linewidth]{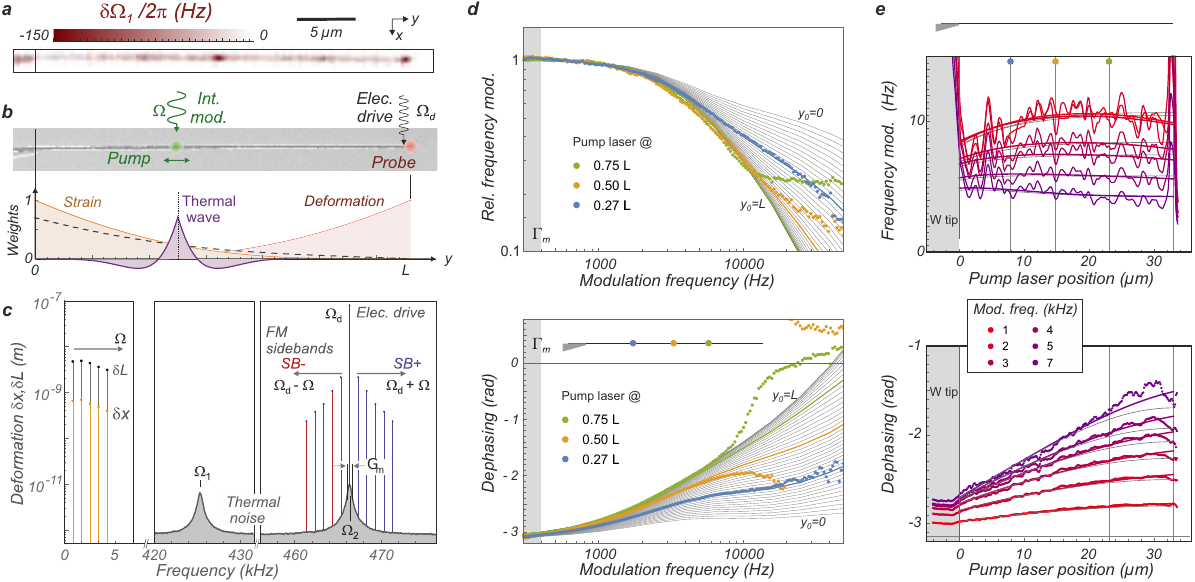}
\caption{
\textbf{Photoinduced dynamical frequency shifts.} {\bf a,b,c} An electrostatic force is used to resonantly drive the $33\,\rm \mu m$ nanowire at its first mechanical resonances  ($\Omega_{\rm d}\approx \Omega_{\rm m}$) and track the frequency shifts (around 425 and 466 kHz) induced by the pump laser. The map represents the static relative  shifts measured while scanning the pump laser along the nanowire. {\bf b} Modulating the pump laser intensity (at a pulsation $\Omega$) generates a dynamical frequency shift $\delta \Omega_{\rm m}[\Omega]$ whose magnitude and dephasing with respect to the intensity modulation is extracted from the dynamical frequency modulation sidebands  $\delta x[ \Omega_{\rm d} \pm\Omega]$ appearing on each side of the resonant drive, while the direct nanowire deformations $\delta x[\Omega], \delta L[\Omega]$ are simultaneously read out by the probe laser.  Panel {\bf b} illustrates the spatial weights  of the strain and deformation contributions for the fundamental longitudinal modes, which are convoluted with the thermal wave profile (purple) to establish the laser-induced dynamical frequency shifts ( $\Sigma_1^2$ (orange) , $ U_1^2$ (red)  and $\Sigma_1^2-\alpha_L/\alpha_E (4 \Sigma_1^2+U_1^2-1)$ (black)).
{\bf d} Relative amplitude and phase of the dynamical frequency shifts measured at different pump laser positions along the nanowire. The grey lines represent the pure photothermal response contributions (strain and deformation only, no force gradient) for different pump laser positions (the colored curves are the ones expected at the respective measurement locations). 
{\bf e} Dynamical frequency shifts measured while scanning the pump laser along the nanowire, for different intensity modulation frequencies. The amplitude measurements are affected by the variability of the local absorption, but not the phase measurements. The grey curves represent the photothermal contributions to the dynamical frequency shifts, while the colored curves include the contribution from the force field gradients induced by the pump laser, which become more visible at the nanowire vibrating extremity  and at large modulation frequencies, when the photothermal contribution vanishes (see text).
}
\label{Fig7}
\end{center}
\end{figure*}
\vskip 1cm
\begin{flushleft}
{\bf Photothermally induced dynamical frequency shifts \\}
\end{flushleft}

Quantitative information on the photothermal response of the nanowire can also be extracted from the analysis of the mechanical frequency changes induced by laser heating. 
Such an investigation is also essential to assess the impact of the probe laser on the frequency stability of those ultrasensitive force  and force gradient probes, which sets their ultimate performances \cite{Fogliano2021}. An electrostatic actuation using a nearby electrode is employed to resonantly drive the nanowire at a frequency $\Omega_{\rm d}/2\pi$, close to one of its mechanical resonances $\Omega_{\rm m}/2\pi$ and  a phase lock loop is implemented to track the laser-induced frequency shifts. The static shifts are shown in Fig.\,7a, obtained  when scanning the pump laser in the nanowire vertical plane. Large absorption spots are clearly visible, at the same positions as in the light-induced deformation maps of Fig 4e, as well as a larger contribution in the clamping area. \\
When the nanowire temperature increases, it undergoes dimensional changes through the thermal expansion coefficient $\alpha_L$ while its stiffness also gets reduced (with a relative strength of $\alpha_E=-3.9\times 10^{-5}\,\rm K^{-1}$). While both mechanisms generate frequency changes, the contribution of the latter is generally ten times stronger than the former. The pump laser induces a non-homogeneous temperature increase within the nanowire which has to be taken into account in the light-induced frequency shifts analyses, as studied in \cite{Pottier2021} in the static regime. In our case, we will time-modulate the pump laser intensity, causing a dynamical modulation of the nanowire  vibration frequencies, to investigate the contribution of the thermal waves to the dynamical frequency shifts. This method can provide a refined understanding of the internal coupling between the eigenmode strain and deformation profiles to the light-induced thermal wave. When the intensity modulation tone is faster than the mechanical linewidth $\Omega > \Gamma_{\rm m}$, the nanowire cannot follow adiabatically the induced frequency shift and, instead, one observes characteristic frequency modulation sidebands, as shown in Fig\, 7c. Here we explain and analyse how to reconstruct the dynamical frequency shifts from those sidebands, in amplitude and in phase, and then compare our measurement to a model describing the interplay between thermal waves and internal deformations of the nanowire. \\

The eigenmodes of the nanowire, of longitudinal spatial profile $u_{\rm m}(y)$ (and curvature $\sigma_{\rm m} (y)\equiv\partial^2_{yy} u_{\rm m}$)  are derived from the Bernouilli equation ( see SI)  and the associated eigenfrequencies $\Omega_{\rm m}$ are settled by: 
\begin{equation}
\frac{1}{2}\int_0^L{\rho S\,  \Omega_{\rm m}^2  u_m(y)^2 dy}= \frac{1}{2}\int_0^L{E I \left(\frac{\partial^2 u_m}{\partial y^2}\right)^2 dy},
\label{eq:omegam}
\end{equation}
where $\rho$ is the nanowire density, $S=\pi R^2$ its surface and $I= \pi R^4/4$ its moment of inertia. When the nanowire temperature locally increases, the local variations of its Young modulus $E(y)=E(1+\alpha_E \delta T(y))$ and dimensions $R(y)=R(1+\alpha_L \delta T (y))$ will modify its eigenfrequencies.  The above expression can be expanded at first order in the temperature change to derive an analytical expression for the light-induced frequency shift. This reasoning also holds in the case of a dynamical modulation of the temperature inside the nanowire, at the condition that the modulation frequency  and its amplitude remains small compared to the mechanical resonance frequency ($\Omega, \delta \Omega_{\rm m} \ll \Omega_{\rm m}$), giving the expression (see SI):

\begin{equation}
\frac{d\Omega_{\rm m}[\Omega]}{\Omega_{\rm m}}
=
 \frac{2}{L} \int_0^L  {\left( \alpha_E \Sigma_m^2 +
\alpha_l \left( 4 \Sigma_m^2 + U_m^2 - 1\right) \right)\delta T [\Omega] dy},
\label{eq.dWmdT}
\end{equation}
where we have introduced the normalized deformation  and strain longitudinal profiles of the mechanical mode: $U_{\rm m} \equiv u_{\rm m}(y)/u_{\rm m} (L)$ and  $\Sigma_{\rm m}\equiv \sigma_{\rm m}(y)/\sigma_{\rm m}(0)$ which are shown in Fig.\,7b. The first term in the parenthesis accounts for the thermally-induced Young modulus changes which only affect the mechanical frequency if they occur in a region of large strain, where the bending energy is large. The second group accounts for the contribution of the geometrical changes. Among them,  the first term reflects the fact  that a nanowire diameter increase in a region of large strain  increases its stiffness and thus its vibration frequency, while the last term conveys that an elongation of the nanowire  causes a frequency softening. This convolution of the modulated thermal wave with the  local strain and deformation profiles, generates a peculiar response function, which strongly varies with the pump laser position and the modulation frequency: at slow modulation, the thermal wave propagates all along the nanowire, while it localizes around the pump laser at larger modulation frequencies, allowing to reveal its interplay with the strain profile $\Sigma_{\rm m}$ which is the dominant contribution in this equation. In addition to those photo-thermally induced dynamical frequency shifts, the pump laser can also generate optical or photothermal force gradients, which also generate dynamical frequency shifts when positioned in a region of large deformation. We note that this ideal model does not taken into account the deformations in the tungsten tip, but this reasoning can be adapted to the entire deformation profile of the suspended nanowire.\\ 

The dynamical frequency modulation $\delta\Omega_{\rm m}[\Omega]$ can be extracted from the relative strength  of the frequency modulation sidebands $\delta x[\Omega_{\rm d}\pm \Omega]$ with respect to the central tone $\delta x[\Omega_{\rm d}]$. A single flexural mode dynamics is described by
\begin{equation}
\delta \ddot x = -\Omega_{\rm m}^2(t) \, \delta x -\Gamma_{\rm m}\, \delta \dot x +\frac{\delta F_d}{M_{\rm eff}} \cos(\Omega_{\rm d}t),
\end{equation}
with $$\Omega_{\rm m}(t)= \Omega_{\rm m} + \delta\Omega_{\rm m} \cos(\Omega t -\phi_m),$$ 
where the amplitude and dephasing of the frequency modulation are described by the complex modulation strength $\delta\Omega_{\rm m}[\Omega]\equiv\delta \Omega_{\rm m} e^{i\phi_{\rm m}}$. For small modulation depths,  ( $| \delta\Omega_{\rm m} | <\Gamma_{\rm m}  $), the relative sidebands strengths are given by
\begin{equation}
\frac{\delta x[\Omega_{\rm d}\pm\Omega]}{\delta x[\Omega_{\rm d}]}\approx \frac{\chi[\Omega_{\rm d}\pm \Omega]}{\chi[\Omega_{\rm m}]}
\frac{\delta\Omega_{\rm m}e^{\pm i\phi_{\rm m}}}{\Gamma_{\rm m}}\underset{{\Omega_{\rm d}=\Omega_{\rm m}}}{\rightarrow}
\frac{\delta\Omega_{\rm m}e^{\pm i\phi_{\rm m}}}{\Gamma_{\rm m}/2-i\Omega}, 
\end{equation}
The validity of this approximation is discussed in the SI. Each lateral sideband thus provides a separate determination of the dynamical frequency modulation strength, while the exploitation of both measurements helps canceling parasitic effects, such as a non-perfect resonant drive or non-linear readout (which intermix the driven displacements at $\Omega_{\rm d}$ and $\Omega$).\\

We first realized  response measurements of the dynamical frequency modulation at three different pump laser  positions $y_0$ while sweeping the modulation frequency $\Omega/2\pi$. The normalized amplitude and dephasing of the frequency modulation with respect to the intensity modulation $\delta \Omega_{\rm m}[\Omega]/\delta \Omega_{\rm m}[0]$ are shown in Fig.\, 7d. The grey curves are the theoretical predictions derived from  expression\, (\ref{eq.dWmdT})  and a clamping contribution (see below) using the same temperature profile as used in the deformation measurements analysis of Fig.\,4, previously performed  on the same nanowire (given by eq.\, (\ref{eq.dT-Rc})). The colored curves correspond to the ones expected  at the  three different measurement locations. One observes an excellent agreement with the experiment for the blue position (at $y_0=0.27 L$) and generally at low frequencies, while a deviation becomes apparent  at larger frequencies, especially for the green position, taken far from the clamping point. This is due to the force gradient contribution, which is more important when the pump laser is positioned at the vibrating extremity of the nanowire. The negative response observed at small modulation frequencies ($-\pi$ dephasing) is in agreement with the frequency softening measured in the static map shown in Fig.\,7a.\\

We then impose a fixed modulation frequency, ranging from 1\,kHz to 7\,kHz, scan the pump laser position along the nanowire and report the dynamical frequency modulation strength as a function of the pump laser position. The amplitude signals are altered by the non-homogeneous absorption of the nanowire, but this was partially corrected by normalizing the response by the lateral response measurements at low frequency (see SI). The phase measurements are not affected by this absorption variability, and are in good agreement with the theoretical predictions shown as full lines: increasing the modulation frequency, one observes a progressive delay arising in the response, due to the propagation time  of the thermal wave from the laser source to the area of large dynamical strain for the first longitudinal mode family, located in the left part of the nanowire. The grey curves  account for the photothermal frequency shifts alone, with no free parameter. We included a tip contribution proportional to the temperature increase at the clamping point $\left.\frac{d\Omega_{\rm m}}{\Omega_{\rm m}}\right|_{\rm tip}= \mu \alpha_E^W \Sigma_{m}^2(0)\, \delta T[0^-,\Omega]$, with $\mu=0.39$ which helps obtain a better agreement with the  measurements, especially close to the clamping area.\\ 
This added term can be viewed as a patch to the model, which does not take into account the strain modulation in the clamping area: a vibrating nanowire also deforms the tungsten tip, and creates a large modulation of its internal stress, which  is at the origin of the losses in the supporting material \cite{WilsonRae2008,Anetsberger2008}.  The temperature sensitivity of tungsten Young modulus is larger than the one of SiC ($\alpha_E^W = 4.8\, \alpha_E$), and this effect represents a non-negligible contribution to the dynamical frequency shifts observed in our system, even detectable when the heat source remains located in the nanowire. The individual contributions are shown in the SI. In parallel, a large negative dynamical frequency shift is also observed when the pump laser is located on the tip end (see Fig.\,7e, grayed area), which underlines  the presence of a dynamical strain modulation in the tip. In our case, the clamping conditions are not sufficiently well controlled to extract meaningful information on the clamping beyond what we have proposed,  but these observations highlight the potential of this novel dynamical sensing methodology to obtain subtle information on the internal deformation properties of the nanomechanical oscillator. \\
The colored curves further incorporate force field gradient contributions from radiation pressure  $\left.\frac{d\Omega_{\rm m}}{\Omega_{\rm m}}\right|_{\rm rad}\propto  U_{\rm m}^2(y_0) $  and cross section modulation forces  $\left.\frac{d\Omega_{\rm m}}{\Omega_{\rm m}}\right|_{\rm cs}\propto  U_{\rm m}^2(y_0) \delta T[y_0,\Omega]$, and provide a better qualitative agreement with the measurements in the vibrating extremity area, especially at larger modulation frequencies, when the photothermal contributions are less efficient.\\
Those measurements are thus in good agreement with the dynamical deformation analysis and complement the study of the photothermal response of the nanowires. They provide a direct connection to the dynamical strain profile within the nanowire  and to the different other mechanisms impacting the mechanical frequency  of the nanowire and thus its stability. This novel methodology is thereby very interesting to implement so as to optimize the operation of those ultrasensitive force probes and mitigate the impact of the probe light field which becomes an essential task especially at low temperatures \cite{Fogliano2021a}.
\\

\begin{flushleft}
{\bf Conclusions \\}
\end{flushleft}

We have  presented different methods and measurement techniques that help understand and characterize the photothermal response of our suspended silicon carbide nanowires. Exploiting the unique opportunity to scan the time-modulated pump and probe lasers along the nanoresonator length, we investigated the nanowire response in the spatial and temporal domains, thus granting access to the heat propagation properties within the nanowire. Elongation measurements were used as a reliable measurement of their absorption and diffusion properties. Tuning the pump laser color allowed  revealing the strong impact on its absorption properties of the internal Mie resonances, as well as on the cross section modulation force induced by the pump laser. The temperature dependence of the nanowire cross section and thus of its optical reflectivity  also served to directly image the thermal wave within the nanowire, allowing a direct evaluation of its internal diffusion coefficient, unaffected by the clamping imperfections, while their thermal contact resistance could be retrieved from a spatial analysis of their photothermal response. We then turned to the analysis of the dynamical mechanical frequency modulation induced by the thermal waves. By varying the pump laser position and modulation frequency, we could evaluate the impact of the thermal waves on the definition of the oscillator frequency, and retrieve in particular their interplay with the strain  and deformation profiles, using the same thermal parameters as the ones used in the analysis of the photothermal deformations.\\
As shown in the SI, similar response measurement were realized at lower temperatures. In those preliminary measurements, we observed a slight acceleration of the photothermal response time constant, lower than anticipated using the evolution of bulk values, which points towards an increase of the thermal contact resistance. Furthermore, all the measurements exposed in the main text are realized in the linear response regime and do not depend on the mean optical power employed. This is not the case anymore at low temperatures which points towards a very rapid variation of the nanowire thermal conductance,  as observed in the optical heating curves of ref.\cite{Fogliano2021a}. This will be the subject of future investigations.\\

Those methods are thus extremely powerful to understand the interaction of the nanowire with the incoming light field. They have been implemented at room and moderately cold temperatures, where the heat propagation is dominated by  diffusion, but should still be pertinent at lower temperatures, where the investigation of heat propagation is a subject of strong fundamental interest. They could also be directly implemented on a large variety of nanomechanical oscillators, such as nanotubes or 2D material membranes, which present very peculiar heat transport and optomechanical properties. \\

\begin{flushleft}
{\bf Acknowledgments \\}
\end{flushleft}
C.G. acknowledges PhD funding from a CDSN of Ecole Normale Supérieure de Lyon.  C.D.  and L.J. acknowledge PhD funding from the Program QuanTEdu-France ANR-22-CMAS-0001 France 2030. A.R.,  M.C. and the project were supported by the French National Research Agency (SinPhoCOM project) and by the European Research Council under the EU's Horizon 2020 research and innovation program, grant agreements No 820033 (AttoZepto).\\

\newpage

\section*{Appendix}

\appendix

\section{Radiation  and conduction losses}

Exploiting the Stefan-Boltzmann law, the total power $P_r$, emitted by  the nanowire surface can be expressed as 
\begin{equation}
  P_r = \int_0^L \epsilon \sigma_S\,T^4\, 2\pi R dy,
\end{equation}
where $\sigma_S = 5.67\times 10^{-8}\, \rm W/m^{2}/K^{4}$ is the Stefan-Boltzmann constant and $\epsilon$ the black body coefficient. It amounts to $ P_r^0=  \sigma T_0^4 2\pi R L\approx 30\,\rm nW$ for a $ 100\,\rm \mu m$ long, 100\,nm radius nanowire thermalized at room temperature $T_0= 300\,\rm K$, knowing that at equilibrium, the nanowire receives a similar amount from the environment. 
When the laser is positioned at the nanowire extremity, the static temperature profile  $T(y)= T_0+\Delta T\,  y/L$  leads to a modification of the emitted power to
\begin{equation}
  P_r =  P_r^0 \frac{T_0}{ 5\Delta T}\left(\left(1+\frac{\Delta T}{T_0}\right)^5-1 \right) \approx P_r^0 \left(1+ 2\frac{\Delta T}{T_0}\right),
\end{equation}
the last expression being obtained for a small temperature increase $\Delta T\ll T_0$.  As such, the radiated power change then amounts to $2 P_r^0 \Delta T/T_0 \approx 200\,\rm pW $, for a 1\,K temperature increase, which is produced by a 100\,nW absorbed power for this nanowire geometry (thermal resistance of $R_{\rm nw}=\frac{L}{\kappa \pi R^2} = 8.8\times 10^6\,\rm K/W$). This 1/500 ratio suggests that the radiative heat flux can be neglected in our case.\\

To study the impact of convection losses, caused by the surrounding air channeling heat away from the nanowire, we conducted elongation measurements at different pressures in the chamber, from $ 10^{-6}\,\rm mbar$ to 1000\,\rm mbar. We observed a strong attenuation of the driven elongation at larger pressures, but also an absence of evolution for pressures lower than 0.1\,mbar in the chamber. Most of the measurements shown here are taken at static pressures lower than 0.01 mbar, so we also neglected this contribution.

\section{Heat diffusion in the nanowire}
\subsection{Dynamical temperature profile}
We solve the driven 1D heat diffusion equation in the nanowire and  give the analytical expressions of the dynamical temperature profile as well as of the dynamical elongation, in absence and presence of a thermal contact resistance, and for two different heat source spatial profiles ( Gaussian or Dirac ). 

As introduced in the manuscript, the temperature evolution along the nanowire $T(y,t)$  in presence of a time modulated heat source follows a 1D heat propagation equation given by:
\begin{equation}
\rho C\, \partial_t T + \partial_y( - \kappa \partial_y T ) = A_{\rm abs} v_0(y) \left ( P_0 + \delta P_0 \cos \Omega t \right).
\end{equation}
The heat source spatial density profile employed will be given by $v_0(y)= \ \delta (y-y_0)/\pi R^2$ or by  $v_0(y)= \sqrt{\frac{2}{\pi w_0^2}} \frac{e^{- 2 (y-y_0)^2/w_0^2}}{\pi  R^2}$, where $y_0$ is the pump laser position and $w_0$ the laser waist size. The Dirac formulation  leading to simpler expressions, it will be employed when one does not operate at too large modulation frequencies (thermal wavelength smaller than the waist size), or when one investigates the  physics close to the waist area. Both results are identical when the waist size shrinks to zero, and we will only use the Gaussian formulation in the direct imaging of the thermal waves. 
The solution to this linear equation can be separated between a static  and dynamical terms:
$$T(y,t)= T_{\rm stat} (y)+ {\rm Re} \left(\delta T[y,\Omega] e^{-i\Omega t} \right)$$
where we have introduced the complex representation of the driven solution.\\

When needed, in the following we also used the dependency  $\delta T[y,y_0,\Omega]$, which stands for the complex representation of the thermal modulation occurring at position $y$ for a driving pulsation $\Omega$ and for a pump position $y_0$. The latter dependency being omitted when non-necessary ($\delta T[y, \Omega]$).\\

The boundary conditions correspond to an absence of energy loss at the vibrating tip $\partial_y T (L,t)=0$ and to either an ideal reservoir at the clamp $T(0,t)=T_0$ or to a contact resistance:  
$ T(0,t)-T_0= - R_c j_{\rm th}(0,t) S = \left.\kappa R_c S\partial_y T\right|_0 $  (with no thermal inertia here). \\
For the punctual heat source, the static temperature profile is:
$$T_{\rm stat}(y)= T_0 + R_c P_{\rm abs}+ \frac{P_{\rm abs}}{\kappa S} \left( y+ \Theta(y-y_0) (y_0-y) \right). $$
The dynamical part is solution of:
\begin{equation}
\left(\partial^2_{yy} +\frac{i \Omega}{D}\right)  {\delta T[\Omega,y]} = -\frac{\delta P_{\rm abs}}{ \kappa S} \ \delta (y-y_0),
\end{equation}
and the boundary condition becomes:
$$ \delta T [0,\Omega] = R_c\, S \kappa \partial_y\delta T[0,\Omega] $$

If we introduce $k\equiv \sqrt{-i\Omega/D}$, the generic solution of the differential equation is:
$$\delta T= A e^{k y}+ B e^{-k y} + \frac{\delta P_{\rm abs}}{2 \kappa S\,  k} \, \Theta(y-y_0)\left( e^{k (y_0-y)} - e^{k (y-y_0)}\right)$$

The two boundary conditions can then be written:
$$A e^{k L}-Be^{-k L}=\frac{s_0}{2k}\left( e^{k(y_0-L)}+e^{k(L-y_0)}\right) $$
$$
A+B = R_c \kappa S k (A-B)
$$

so that the solution is

\begin{widetext}
\begin{equation}
\frac{\delta T [y,\Omega]}{ \delta P_{\rm abs}}=
\frac{R_{\rm nw} }{2k L}\left(
\frac{e^{k (L-y_0)}+e^{k (y_0-L)}}
{e^{kL}(1+ k L_c)+e^{-k L} (1-k L_c)}
\left((1+k L_c) e^{k y}-(1-k L_c) e^{-k y}\right)
+\Theta(y-y_0) (
e^{k (y_0-y)}-e^{k (y-y_0)}
)
\right)
\label{eq.dT-Rc}
\end{equation}
\end{widetext}
where we have introduced the nanowire thermal resistance $R_{\rm  nw}= L/\kappa $, and the contact resistance equivalent length $L_c= L R_c/R_{\rm nw}$.  This expression equals to eq. (2) of the manuscript in absence of  contact resistance.  One can verify that it converges towards the expression of $T_{\rm stat}(y)-T_0$ when the modulation frequency  approaches 0 ($\Omega, k \rightarrow 0$).\\

In the configuration of thermal wave imaging,  we operate at large modulation frequencies, at positions $y, y_0$ in the middle of the nanowire, such that the thermal wavelength becomes  sufficiently short to only keep the dominant exponential terms in the previous expression, which then simplifies to:
\begin{widetext}
\begin{equation}
\frac{\delta T [y,\Omega]}{ \delta P_{\rm abs}}=
\frac{R_{\rm nw} }{2k L}\left(
\frac{e^{k (L-y_0)}}
{e^{kL}(1+ k L_c)}
\left((1+k L_c) e^{k y}\right)
+\Theta(y-y_0) (
e^{k (y_0-y)}-e^{k (y-y_0)}
)
\right) 
= 
\frac{R_{\rm nw} }{2k L}e^{k |y-y_0|}.
\label{eq.dT-Rclocal}
\end{equation}
\end{widetext}
From this, we retrieve the $k^{-1}\propto \Omega^{-1/2}$ evolution observed at large frequencies, and the corresponding $+\pi/4$ dephasing. 

\subsection{Elongation}
The static elongation of the nanowire  is given by:
$$\Delta L _{\rm stat}= \int_0^L \alpha_L (T_{\rm stat }(y)-T_0)\, dy $$

and amounts to:
\begin{widetext}
$$\frac{\Delta L_{\rm stat}}{L} =\alpha_L P_{\rm abs}\left(R_c+ R_{\rm nw}\left(\frac{y_0}{L}-\frac{y_0^2}{2 L^2}\right)\right) $$
\end{widetext}
while the dynamical elongation $\delta L _{\rm stat}[\Omega]= \int_0^L \alpha_L \delta T[\Omega,y] dy $ is given by

\begin{widetext}
\begin{equation}
\frac{\delta L[\Omega]/L}{\delta P_{\rm abs}}=\frac{\alpha_L R_{\rm nw}}{2 k^2 L^2} \left(
\frac{e^{k  (y_0-  L)}+ e^{k ( L- y_0)}}
{e^{kL}(1+ k L_c)+e^{-k L} (1-k L_c)}
\left((1+kL_c)e^{k L}+(1-kL_c)e^{-k L}-2\right)
+
2- e^{k ( y_0- L)}- e^{k ( L- y_0)}
\right).
\label{eq.dL-Rc}
\end{equation}
\end{widetext}
When the laser is positioned at the vibrating extremity of the nanowire, it becomes :

\begin{widetext}
\begin{equation}
\frac{\delta L[\Omega]/L}{\delta P_{\rm abs}}=\frac{\alpha_L R_{\rm nw}}{ k^2 L^2} \left( 1-
\frac{2}
{e^{kL}(1+ k L_c)+e^{-k L} (1-k L_c)}
\right).
\label{eq.dL-Rc2}
\end{equation}
\end{widetext}
and when expanded at second order in $kL$, $k L_c$ (first order in $\Omega/\Omega_c $) we have
\begin{equation}
\frac{\delta L[\Omega]}{L }
=
\frac{ \alpha_L R_{\rm nw} }{2 }(1 + 2 L_c/L )\frac{\delta P_{\rm abs}}
{1-i\Omega\frac{L^2}{2 D}(1 + 2 L_c/L )}.
\end{equation} 
 We then extract the thermal cutoff and the effective diffusion coefficient $D/(1+2L_c/L)$.

\section{Absorption estimation}

We compute the absorbed optical power from the analytical expressions of the  electromagnetic fields given by the Mie theory applied to an infinite homogeneous cylinder \cite{Bohren1983}.
The  total absorbed power $w_a$ is  determined from the integral of the Poynting vector $\mathbf{S}$, over a closed surface surrounding the nanowire:
\begin{equation}
    w_{\rm a} = \oint_{\mathcal{A}} \mathbf{S}\cdot {\rm d}\boldsymbol{\sigma}.
\end{equation}
In absence of material absorption, this quantity is zero, all the incoming light being completely scattered. The time-averaged Poynting vector field around the nanowire can be expressed as a function of the incident and scattered fields,
\begin{equation}
    \mathbf{S} = \frac{1}{2\mu_0}\text{Re}\left[ (\mathbf{E}_{\rm inc} + \mathbf{E}_{\rm sca}) \times (\mathbf{B}_{\rm inc} + \mathbf{B}_{\rm sca})^* \right].
\end{equation}

The problem is analytical in the case of an incident plane wave illumination. Since the nanowire is surrounded by vacuum, we can extend the closed surface of integration to regions far from the nanowire surface. We can thus exploit asymptotic expressions of the fields to optimize the numerical calculation.
To refine our evaluation, we expand the focussed incident field on a set of incoming plane waves featuring different angles of incidence \cite{Reigue2023} allowing us to account for the focusing effects of the microscope objectives employed and for the variations with the selected wavelength of the AOTF-filtered super-continuum pump beam size.

\begin{figure}[t!]
\begin{center}
\includegraphics[width=0.95\linewidth]{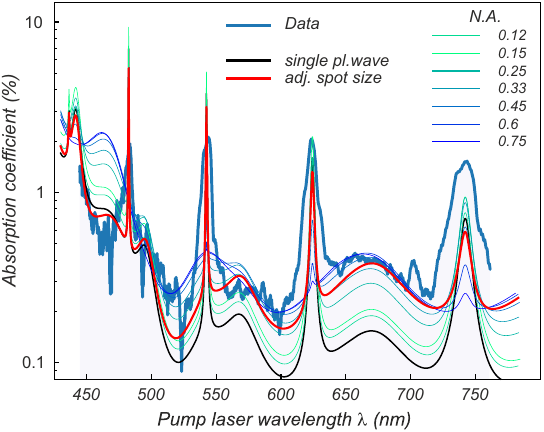}

\caption{
Comparison between the measured spectral structure of the absorption coefficient (thick blue line) and our model absorption.
We plot the computed absorption coefficient $W_a/P_{0}$, with $P_0$ the incident optical power, when considering a single plane wave illumination (thick black line), and for a 2D Gaussian beam shaped incident field with various NA objectives (coloured thin lines).
In red, the computed absorbed coefficient when accounting for the variation of the focused laser waist with the Super-continuum output colour (the output waist prior to the focussing objective (100$\times$ 0.75\,NA) in the blue is smaller than in the red).
}
\label{Fig_resonant_absorption}
\end{center}
\end{figure}

The absorption of the incident optical power occurs within the nanowire cross section due to the bulk absorption of our 3C-SiC nanowires.
The imaginary part of the refractive index is set so that it follows the penetration depth $a$ found in  literature \cite{Patrick1969}, using
\begin{equation}
    \text{Im}(n) = \frac{a}{k}, \ \  {\rm with}\ \ 
    a = K\,\left(E_{\rm ph} - E_{\rm g}\right)^2 \,\frac{1}{1+e^{\frac{E_{\rm ph} - E_{\rm g}}{k_{\rm B}\,T}}} + a_{\rm bck},
\end{equation}
where $k$ is the optical wavevector, $E_{\rm ph}$  the photon energy, $E_{\rm g}$  the indirect bandgap energy of 3C-SiC, $k_{\rm B}$ the Boltzmann constant, $T$ the temperature, and $K= 6.2\times10^{40}$\,cm$^{-1}$J$^{-2}$ the constant used to fit the measured data.
We also added a constant absorption background at $a_{\rm bck}= 40\,\rm cm^{-1}$ to account for the residual absorption observed beyond the gap, although one could employed a more refined function to better fit the data.
This modified refractive index  is inserted in the Mie expressions of the fields to compute the total EM fields surrounding the nanowire and evaluate the absorbed power for each wavelength.
In the figure, we have shown the influence of focusing objective's NA, and accounted for variations of the beam size at the different wavelengths (the output size of the super-continuum  collimated  beam  changes with the wavelength, see Fig. \ref{Fig_resonant_absorption}).
We also convolved the predicted response with a square function to account for the finite  spectral selectivity of the AOTF filter (0.5-1\,nm over the used color range). We obtain a good qualitative agreement with the measurements. The positions of the Mie resonance are precisely reproduced, but their measured linewidths are always a bit larger than theoretically expected, corresponding to a lower quality factor. This is likely an effect of the nanowire surface rugosity and geometrical imperfections.

\section{Imaging the photothermal wave}

The heat waves generate local changes in the nanowire optical reflectivity, mainly caused by the temperature dependence of their refractive index,   which we track through our optical measurement scheme. We have shown that by probing such an effect at high modulation frequency, we can directly image a thermal wave propagating around a given heat source point. In our experiment, we generate heat waves with the pump laser (532\,nm) and probe the reflectivity changes induced in the probe laser's wavelength ($633\,\rm nm$). 

\begin{figure}[t!]
    \centering
    \includegraphics[width=0.99\linewidth]{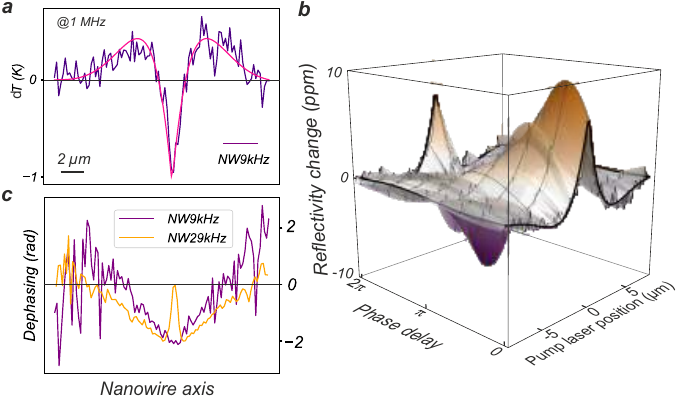}
    \caption{\textbf{Thermal wave imaging. a.} We image a diffusive thermal wave following the principle expounded in the main body of the article. In the case of this nanowire, the reflectivity changes are dominated by thermal effects and our 1D-diffusion model suffices to fit the measurement. As a result we find a local temperature increase of the order of $1\,K$ where the two lasers superimpose, in agreement with theoretical predictions extrapolated using this nanowire's (NW9kHz) thermal diffusion coefficient. \textbf{b.} By varying the representation time ($t$) over one temporal period of the drive, we can picture the evolution of its propagation. Theoretical curves derived from the fitted parameters are displayed alone for clarity. \textbf{c.} We compare the dephasing of reflectivity changes with respect to the optical excitation measured on two different nanowires. NW29kHz (shown in the manuscript) presents a  localized instantaneous response when both lasers are super-imposed, which  is not the case for NW9kHz where thermal effects are dominant. This local effect is a signature of electrostriction (light-induced strain in the material).}
    \label{fig:EL}
\end{figure}

The simplest case we have studied is that of NW9kHz, shown in Fig.\ref{fig:EL}-a. As described in the main body of the article, we retrieve the local temperature changes through the thermo-optical constant of cubic SiC and plot the real part of the response ${\rm Re} \left(\delta T[y_{\rm probe}, y_0,\Omega] e^{-i\Omega t} \right)$ resulting in the profile of the diffusive thermal wave (shown at $t=0 $ here). The theoretical predictions given by the temperature variations around the probe laser's position and their dependency on the modulation frequency show a solid agreement with the measurements, from which we derive the internal thermal diffusion coefficient of the nanowire. It is possible  to  represent the propagation of the driven thermal wave by varying the displayed phase ($\Omega t$) over one period as shown in Fig.\ref{fig:EL}-b, where the only the theoretical profiles derived from the fit are shown alone for clarity's sake. The critically damped nature of the thermal wave can be well identified.

The data shown in the article (using NW29kHz) presents an additional contribution to the a simple thermal wave: the dephasing map shows a peculiar central peal where the two lasers superimpose (see Fig.\ref{fig:EL}-c), which  instantaneously follows the intensity modulation (0 dephasing). This structure is not present on the other nanowire (nor is it expected by our thermal diffusion model) hinting at an additional contribution to light-induced reflectivity changes. The measured curve can  be well fitted in amplitude and in phase when including an instantaneous Gaussian-shaped contribution (when both lasers are focused on the nanowire), using the amplitude and phase of the complex response:
\begin{equation}
    \delta R_{\rm opt} [\Omega,y_0] =  A \,  \delta T[y_{\rm probe}, y_0,\Omega] + B e^{-2\frac{(y_{\rm probe}-y_0)^2}{w_0^2}},
\end{equation}
where $w_0$ is the laser waist dimension, and A and B the weights of each term. We verified that this local effect was not an artefact (modulated green light collected onto the red photodetectors) by adding additional filters. This extra contribution could originate from a Kerr effect or from electrostriction. The former relates $\Delta n$ to the optical power input via the Kerr constant $n_2 \sim 10^{-19}\,\rm m^2/W$ \cite{DeLeonardis2017} which results in $\Delta n_{\rm Kerr} = 10^{-13}$, which is largely negligible compared to the measured $\Delta n_{th} \approx 10^{-5}$ \cite{Powell2020}. Electrostriction effect originates from the reflectivity changes induced by the mechanical stress induced in the nanowire by the modulated pump light (through the photoelastic constant $p \sim 10^{-1}$ \cite{Djemia2013}). For a 1\,mW beam, the stress modulation amounts to \cite{Rakich2010} $\delta \sigma^{\rm es}= p \epsilon_0 n^4 E^2 \approx 10^4 \,\rm Pa $, and typically 50 times more if one accounts for the Mie resonance enhancement of the intra-nanowire pump intensity. This would generate a relative strain and reflectivity change of $\delta R_{\rm opt}/R_{\rm opt} \simeq \delta \sigma ^{\rm es}/ E$ of few ppm without accounting for the enhanced sensitivity to the internal refractive index change at the probe wavelength caused by the Mie resonances. \\
In our measurements, we found that NW29kHz presents a Mie resonance in the vicinity of the pump wavelength (532\,nm), contrary to NW9kHz, which could explain the absence of local response from the latter. Additional measurements, realized at different pump wavelengths, could help clarify this statement. However this effect remains spatially localised under the probe laser and does not prevent us from exploiting those measurements to image and  analyze the thermal wave properties, that spread significantly beyond the laser spot.

\section{ Dynamical frequency shifts}

Solving the Bernoulli equation for the lateral deformations of the nanowire:
$$\rho S \frac{\partial^2 u}{\partial t^2}=-\frac{\partial^2}{\partial y^2}\left( E I \frac{\partial^2 u}{\partial y^2}\right)+ f$$
 the eigenmodes profiles  are  given by
 \begin{widetext}
$$u_n(y)
\equiv
\frac{
-(\sin k_n L+ \sinh k_n L)(\cos k_n y-\cosh k_n y)
+
(\cos k_n L +\cosh k_n L)(\sin k_n y-\sinh k_n y)
}{
2 (\sin k_n L \cosh k_n L-\cos k_n L\sinh k_n L)}$$
\end{widetext}
where $ k_1 L\approx 1.875,\,\,\, k_2 L\approx 4.694,\,\,\, k_3 L\approx 7.855 $,
satisfying  $u_n(0)=0$,  $u_n(L)=1$ and $\int_0^L u_n(y)^2 dy= L/4$. The strain profile is proportional to their second spatial derivatives:

\begin{widetext}
$$\sigma_n(y)
\equiv- k_n^2
\frac{
-(\sin k_n L+ \sinh k_n L)(\cos k_n y+\cosh k_n y)
+
(\cos k_n L +\cosh k_n L)(\sin k_n y+\sinh k_n y)
}{
2 (\sin k_n L \cosh k_n L-\cos k_n L\sinh k_n L)}$$
\end{widetext}
which satisfies  $\sigma_n(0)=k_n^2$,  $\sigma_n(L)=0$ and $\int_0^L \sigma_n(y)^2 dy= k_n^4 L/4$. With this definition, the oscillator frequency is given by :

\begin{equation}
\frac{1}{2}\int_0^L{\rho S \Omega_{\rm m}^2 u_m(y)^2 dy}= \frac{1}{2}\int_0^L{E I \left(\frac{\partial^2 u_m}{\partial y^2}\right)^2 dy},
\label{eq:omegam}
\end{equation}
where $\rho= 3200\,\rm kg/m^3$ is the SiC density, $E=400\,\rm GPa$ its Young modulus, $I= \pi R^4/4 $ the quadratic moment of inertia. Using the integrals given above, we obtain:
$$\Omega_{\rm m}^2 =\frac{E I}{\rho S} k_m^4= \frac{E }{4\rho} \frac{R^2}{L^4} (k_m L)^4,$$ 
or $\Omega_{\rm m}/2\pi \approx 0.28\sqrt{ \frac{E}{\rho}} \frac{R}{L^2}$ for the first longitudinal eigenmode. 
\vskip 1cm
\begin{flushleft}
{\bf  Light-induced dynamical frequency shifts,  temperature sensitivity}
\end{flushleft}

We now use a variational method to investigate, at first order, the impact of a temperature variation. It causes a change in the material stiffness (relative coefficient $\alpha_E$ in $\rm K^{-1}$) as well as a thermal expansion governed by the relative linear coefficient $\alpha_l$. The  temperature profile is given by $T(y,t) =  T_0+ \delta T (y,t)$. The material stiffness becomes position dependent:
$$ E= E_0(1+\alpha_E \delta T),$$
where we have introduced the stiffness at the  experiment temperature $E_0$. The material expansion will locally modify its dimensions to:
$$ R= R_0(1+\alpha_l \delta T),$$
$$ \rho S= \rho_0S_0(1-\alpha_l \delta T),$$
$$I=I_0(1+4 \alpha_l \delta T).$$
The total length of the nanowire will be modified according to :
$$L= L_0+\int_0^L{ \alpha_l \delta T dy}.$$
Incorporating those perturbations in  expression (\ref{eq:omegam}),  the frequency shifts $d\Omega_{\rm m}/2\pi$ are defined by:
\begin{equation}
\begin{array}{l}
\int_0^{L+dL}{\rho_0 S_0 (1-\alpha_l\delta T) (\Omega_{\rm m}+d\Omega_{\rm m})^2 u_m(y)^2 dy}\\
= \int_0^{L+dL}{E_0(1+\alpha_E\delta T) I_0(1+4\alpha_l\delta T) \left(\frac{\partial^2 u_m}{\partial y^2}\right)^2 dy},\\
\end{array}
\label{eq:domegam}
\end{equation}
after a bit of simplification, keeping only the first order terms, we have:
\begin{widetext}
\begin{equation}
\frac{d\Omega_{\rm m}}{\Omega_{\rm m}}
=\frac{1}{2}\left(
(\alpha_E+4\alpha_l) \int_0^L{\frac{\sigma_n(y)^2}{\int_0^L{\sigma_n(y)^2 dy}} \delta T(y)dy}
+
\alpha_l \int_0^L{\frac{u_n(y)^2}{\int_0^L{u_n(y)^2 dy}} \delta T(y)dy}
-4 \frac{dL}{L}
\right)
\end{equation}
\end{widetext}
If we introduce the normalized strain and deformation weighting functions, we obtain:
\begin{widetext}
\begin{equation}
\frac{d\Omega_{\rm m}}{\Omega_{\rm m}}
=
\alpha_E  \frac{2}{L} \int_0^L{\frac{\sigma_n(y)^2}{\sigma_n(0)^2} \delta T(y)dy}
+
\alpha_l \frac{2}{L} \int_0^L{\left( 4 \frac{\sigma_n(y)^2}{\sigma_n(0)^2} + \frac{u_n(y)^2}{u_n(L)^2}-1\right) \delta T(y)dy}
\end{equation}
\end{widetext}
Those weighting functions are shown in Figure 7 of the manuscript. The frequency shifts caused by a temperature-induced stiffness change are efficient only if the temperature increase appears at locations of large dynamical strain, in the first part of the nanowire for the fundamental mode.
On the other hand, deformational effects show opposite signs at the two nanowire extremities: a positive frequency shift is observed on the clamped end where an increasing diameter increases the stiffness, while a temperature change inducing a considerable elongation, maximized when heating at the free end, softens the mechanical frequency.
This local behaviour can be observed at sufficiently large modulations frequencies, when the thermal wave remains local.\\
In the case of a time-modulated temperature change, we can express the dynamical frequency shifts as
\begin{widetext}
\begin{equation}
\frac{\delta\Omega_{\rm m}[\Omega]}{\Omega_{\rm m}}
=
\alpha_E  \frac{2}{L} \int_0^L{\frac{\sigma_n(y)^2}{\sigma_n(0)^2} \delta T[y,\Omega]dy}
+
\alpha_l \frac{2}{L} \int_0^L{\left( 4 \frac{\sigma_n(y)^2}{\sigma_n(0)^2} + \frac{u_n(y)^2}{u_n(L)^2}-1\right) \delta T[y,\Omega]dy } + \frac{1}{2 M_{\rm eff} \Omega_{\rm m}^2} \nabla \delta F_{\rm ext}[\Omega] \frac{u_n^2 (y_1)}{u_n^2(L)}
\label{eq.dWm}
\end{equation}
\end{widetext}
where we have added the contribution of an external force field gradient produced by the pump laser located at $y_0$ (see below for the contributions of the different optical and photothermal forces).\\
\begin{figure*}[t!]
\begin{center}
\includegraphics[width=0.9\linewidth]{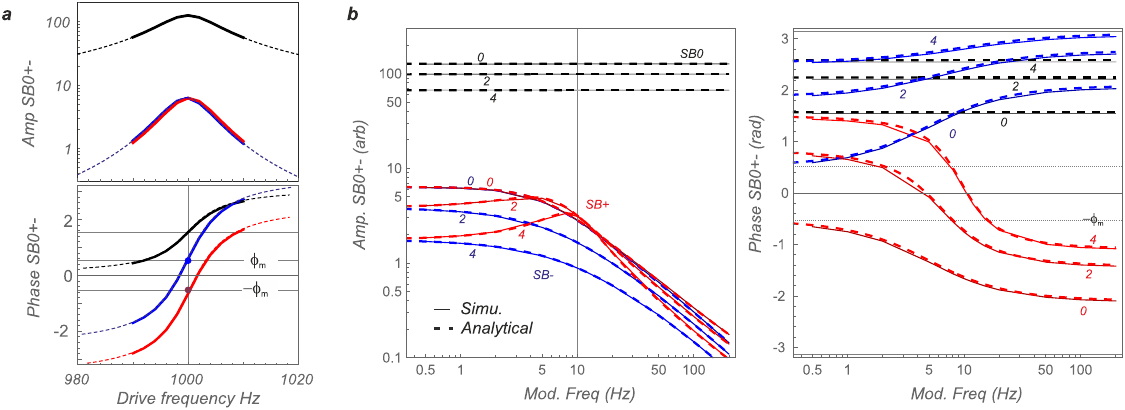}
\caption{
\textbf{Numerical simulations.} Here we simulate the dynamical equation and compare the demodulated sideband strengths to the analytical expressions.  To reduce the computation duration, we simulate here the dynamical equation using a low Q oscillator, with frequency of $\Omega_{\rm m}/2\pi= 1\,\rm kHz$ and damping  $\Gamma_{\rm m}/2 \pi= 10\,\rm  Hz$.  {\bf a Quasi-static response.} Amplitude and dephasing of the main sidebands, measured at  low modulation frequency $(\Omega/2\pi =0.5\,\rm Hz)$ and low modulation depth ($\delta\Omega_{\rm m}/2\pi =0.5\,\rm Hz$ with a dephasing of $\phi_{\rm m}=\pi/6$) for varying detuning of the drive frequency $\Omega_{\rm d}$. While the central sideband (black) follows the resonant mechanical response (governed by $\chi[\Omega_d]$), the lateral sidebands (red, blue at $\Omega_{\rm d}-\Omega$, $\Omega_{\rm d}+\Omega$ respectively) present a different  spectral evolution, associated to $\chi[\Omega_d]^2$ at low modulation frequency. This results in a larger cumulated  dephasing across the mechanical resonance, which amounts to $2\pi$  for an instantaneous  frequency modulation ($\phi_{\rm m}=0 $).  At resonance, one recovers the $\pm \phi_m=\pm \pi/6$ phase, which was arbitrarily  used in those simulations. {\bf b Dynamical response at small modulation depth.} Amplitude and dephasing of the three main sidebands, when one sweeps the modulation frequency  $\Omega/2\pi$ of the mechanical resonance. 
The modulation depth employed here is small (0.5 H)z. and the drive is  detuned by 0, +2 and +4 Hz which creates an unbalance of the sidebands, especially at low modulation frequencies. Numerical simulations are shown using full lines and analytical expressions with dashed lines.
}
\label{fig-S2}
\end{center}
\end{figure*}

In the case of a homogeneous temperature increase $T\rightarrow T+\Delta T$, the mechanical frequency shifts can be expressed as:
$$\frac{\Delta\Omega_{\rm m}}{\Omega_{\rm m}}=\left( \frac{1}{2} \alpha_E +\frac{1}{2}\alpha_l \right) \Delta T$$
where the $1/2$ coefficients are in agreement with the strain and deformation dependencies observed in the eigenfrequency equation.

\vskip 1cm
\begin{flushleft}
{\bf  Extraction of the dynamical frequency shifts from the sideband strength}
\end{flushleft}

We now describe how the dynamical frequency modulation strength $\delta\Omega_{\rm m}[\Omega]$ is extracted in the experiment, from the frequency modulation sidebands. We compared our analytical expressions, valid at low modulation strengths, to numerical simulations.\\
The dynamics of a 1D nanowire, subjected to a driven frequency modulation $ \delta\Omega_{\rm m} \cos\left(\Omega t-\phi_m \right)$  (related to the complex modulation strength $\delta\Omega_m[\Omega]\equiv \delta\Omega_m e^{i\phi_m}$ ) caused by a laser modulation at frequency $\Omega/2\pi$ with a temporal delay $\phi_m[\Omega]/\Omega$ with respect to the power modulation, while being simultaneously driven by a quasi-resonant electrostatic force $\delta F_{\rm d}\cos(\Omega_d t)$ is described by

\begin{equation}
\begin{array}{lll}
\delta  \ddot{x}(t) &=& -\left(
  \Omega_{\rm m} +\delta\Omega_{\rm m} \cos(\Omega t-\phi_m )
  \right)^2\ \ \delta x(t)\\
  & & \\
  & & -\Gamma_{\rm m}\, \delta \dot{x}(t) +\frac{\delta F_{\rm d}}{M_{\rm eff}}\cos(\Omega_d t).\\
  \end{array}
\end{equation}

In the experiments, the signal quadratures are measured by locked-in detectors demodulating the lateral deformation signal $\delta x(t)$ at the resonant mechanical drive $\Omega_{\rm d}$ and on the first lateral frequency modulation sidebands at $\Omega_{\rm d} \pm \Omega$, as shown in Fig.\,7 of the manuscript. If we use the quadrature definitions
$A_X[\Omega] \equiv 1/T \int_0^T{dt A(t) \cos\Omega t}$, $A_Y[\Omega] \equiv 1/T \int_0^T{dt A(t) \sin\Omega t}$, for which the quadratures of a $A_0 \cos[\Omega t-\phi_0]$ signal are $A_X[\Omega]= A_0 \cos\phi_0/2$, $A_Y[\Omega]=  A_0 \sin\phi_0/2$, we obtain and numerically verified that at small modulation depth, the signal quadratures follow

\begin{eqnarray}
  \delta x[\Omega_d + \Omega]&=& \delta x[\Omega_d] 
\ \ \ 
\frac{1}{2}  \frac{\chi[\Omega_d+\Omega]}{\chi[\Omega_{\rm m}]}
 \frac{\delta\Omega_{\rm m} e^{i \phi_{m}}}{\Gamma_{\rm m}/2} \\
\delta x[\Omega_d - \Omega] &=&\delta x[\Omega_d] 
\ \ \ 
\frac{1}{2}  \frac{\chi[\Omega_d-\Omega]}{\chi[\Omega_{\rm m}]}
 \frac{\delta\Omega_{\rm m} e^{-i \phi_{m}}}{\Gamma_{\rm m}/2}
 \label{eq.SBDlinear}
\end{eqnarray}

A filtering effect arises from the susceptibility ratio, which can be approximated at small modulation frequencies $\Omega \ll \Omega_{\rm m}$ in the resonant drive condition $\Omega_{\rm d}=\Omega_{\rm m}$ by:
\begin{equation}
 \frac{\chi[\Omega_m\pm \Omega]}{\chi[\Omega_{\rm m}]}\approx \frac{1}{1\mp i \Omega/(\Gamma_{\rm m}/2)}.\\
\end{equation}
This represents a first order low-pass filter with a cutoff at $\Gamma_{\rm m}/2$, half the mechanical linewidth.\\

\begin{figure}[t]
\begin{center}
\includegraphics[width=0.99\linewidth]{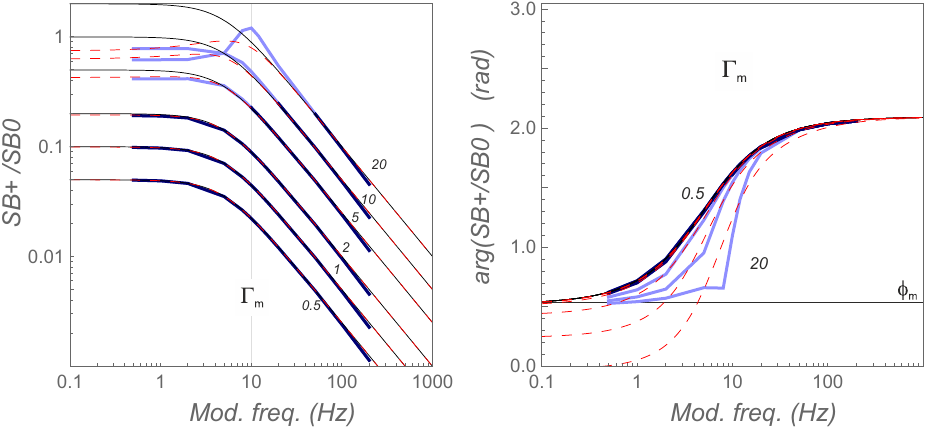}
\caption{
\textbf{Numerical simulations, impact of the modulation depth.} Here, the drive is resonant ($\Omega_{\rm d}=\Omega_{\rm m} $), while we increase the modulation strength : $\delta\Omega_{\rm m}/2\pi= 0.5,1,2,5,10,20\,\rm Hz $.  We plot the relative amplitude and dephasing between the central and upper sidebands $\delta x[\Omega_d+\Omega]/\delta x[\Omega_d]$.
The lighter blue color represents the points for which $\left|\frac{\delta \Omega_{\rm m}}{\Gamma/2-i \Omega}\right| >\frac{1}{2} $, that is when the linear expansion of the Bessel functions is not valid anymore.
The black traces correspond to the linear equations \ref{eq.SBDlinear}, in red the approximated expressions based on Bessel functions $J_1(\frac{\delta \Omega_{\rm m} e^{i \phi_m}}{\sqrt{(\frac{\Gamma_{\rm m}}{2})^2+\delta\Omega_{\rm m}^2} - i \Omega})$. Due to the folding observed at large modulation strengths, one should remain at small modulation strengths so as to univocally convert the measurement signals into the original frequency modulation. This is largely the case in our measurements (even if the folding can still be observed at large modulation strength and low frequency modulation), so the linear expressions can be safely employed in our analyses.
}
\label{fig-S3}
\end{center}
\end{figure}

The numerical verification of those expressions is shown in Fig.\,\ref{fig-S2}and Fig.\,\ref{fig-S3}, where we verified the sideband's dependence on a detuning of the electrosatic drive, on the modulation frequency and on the modulation depth.  The numerical simulations were conducted using a Runge-Kutta 4 algorithm in C. In each simulation run, we first let the system reach a stationary regime for a duration of $500 \Gamma_{\rm m}^{-1}$,before computing the quadratures over 1000 modulation periods. To reduce this simulation time, we employed a lower frequency oscillator, featuring a low Q factor (1 kHz, Q=100). The simulations are in very good agreement with the analytical expressions  as long as :
\begin{equation}
\left| \frac{\delta\Omega_{\rm m} e^{-i \phi_{m}}}{\Gamma_{\rm m}/2-i\Omega} \right|<1,
\end{equation}
shown in darker color in Fig.\,\ref{fig-S3}.
At stronger modulation strengths (ligther points), the simulation deviates from the above equations. 
More advanced simulations allowed us to investigate more complex effects which arise when we take into account the 2D lateral degrees of freedom of the nanowire, where we observed cross-coupling effects.
That being said, in this work we remain at small modulation strengths and modulation frequencies smaller than the 40\,kHz frequency splitting observed  between the two fundamental eigenmodes. Furthermore, we positioned the electrode and the measurement vector to be aligned with the mode of interest  and minimized the actuation efficiency of the other transverse mode, to assimilate our measurements to the 1D case.\\

\begin{flushleft}
{\bf  {Experimental signals}\\}
\end{flushleft}
Dynamical frequency shifts measurements were realized by driving the higher frequency mode of the 440\,kHz nanowire using an electrostatic voltage, with an amplitude of approximatively 100\,nm locked at resonance by a PLL (bandwidth $\simeq 100\,\rm Hz$). The mechanical linewidth is around 400\,Hz, and the driven resonant frequency  460 kHz. The pump laser intensity is  modulated at different frequencies (1,2,3,4,5,7,10\, kHz), and scanned in the vertical plane using a slightly tilted scan plane to maintain the nanowire in the waist.\\ 
The probe laser was positioned at the nanowire extremity, where the difference and sum channels are sensitive to the lateral and longitudinal deformations respectively. For each position within the vertical plane, we recorded the amplitude and the phase of the driven response at  $\Omega_{\rm d}$  and of the frequency modulation sidebands, at frequencies $(\Omega_d\pm \Omega)/2\pi$ on each side of the driving tone, which was locked by the PLL at resonance. We also simultaneously recorded the deformations of the nanowire $\delta r_x[\Omega]$ and $\delta L[\Omega]$, as explained above, as well as the frequency shifts recorded by the two PLLs ($\delta f_1, \delta f_2$) (the first mode being only resonantly driven, without measuring its dynamical sidebands). Each measurement thus represents a spatial XY map, and for each vertical position $y$ we realize a Gaussian fit of the signal amplitude cut along x, extract the adjusted amplitude  and evaluate the phase at the position of maximum amplitude. We thus obtain a measurement along the nanowire length of the different signals: $\delta r_x[\Omega]$, $\delta L[\Omega]$, $\delta r_x[\Omega_d\pm \Omega]$, for the different modulation frequencies $\Omega/2\pi$ employed. \\

\begin{figure*}[t!]
\begin{center}
\includegraphics[width=0.7\linewidth]{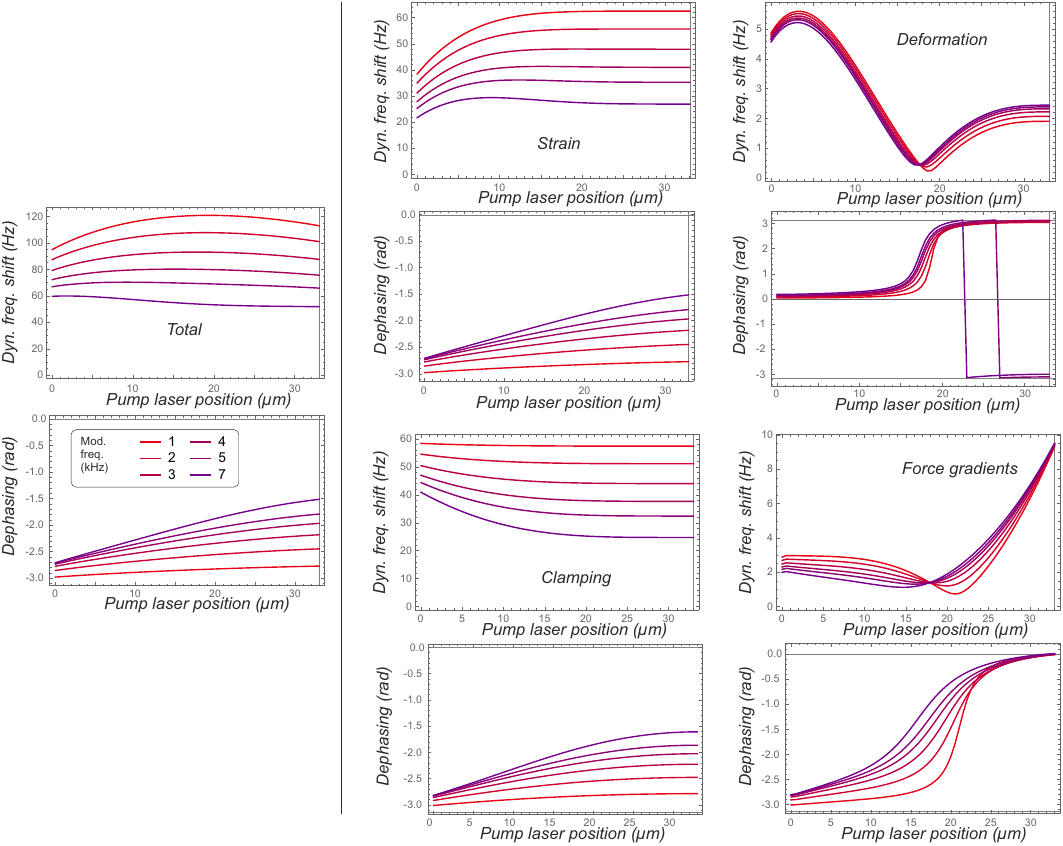}
\caption{{\bf Individual contributions to the  dynamical frequency modulation}
  {\bf Left}: total, {\bf right}:  strain, deformation, tip and force gradient contributions to the dynamical frequency modulation $\delta \Omega_{\rm m}[y_0,\Omega]$. In each plot we report the dependence on the pump laser position $y_0$ of the amplitude and the phase of each contribution, for increasing modulation frequencies, which are used to adjust the experimental results. The grey curves in Fig.\,7e of the manuscript only take into account the first three contributions.
}
\label{Fig_resonant_absorption}
\end{center}
\end{figure*}
The local variations in the absorption coefficient $A(y_0)$ are evaluated  and compensated by exploiting the measurement of the lateral deformations at low frequency, $\delta r_x[\Omega\rightarrow 0]$.  As shown in the manuscript, the low frequency temperature increase  at the clamping point does not depend on the laser position at constant absorbed power, so this measurement can be used to determine the variations of the absorption coefficient.  In practice we used the measurement done at $\Omega_{\rm cal}/2\pi= 1\,\rm kHz$  to evaluate the absorption coefficient which can be extracted using: 

$$\delta r_x[\Omega_{\rm cal},y_0] = \frac{dx}{dT}\,   A (y_0) \delta T[0^+,y_0,\Omega_{\rm cal}] $$
where  the coefficient $\frac{dx}{dT} $   and the dynamical temperature profile are the same ones as employed in the analyses of the nanowire photothermal deformations of Fig.\,4 in the manuscript. This normalization by the  local absorption coefficient helps suppressing part of the response inhomogeneities, but the final amplitude plots still present some irregularities, and most of the quantitative spatial information is better extracted from the phase plots, which are partly insensitive to those absorption irregularities.\\
We then reconstruct the complex dynamical frequency modulation $\delta \Omega_{\rm m}[\Omega]$ experienced by the nanowire using the analytical expressions (\ref{eq.SBDlinear}). To better compensate for the possible lag in the PLL, observed around  the locations of large absorption, where the frequency shifts are larger, at each laser position $y_0$ and optical modulation frequency $\Omega/2\pi$, we average the two measurements realized on each lateral sidebands. \\

This procedure allows extracting the laser-induced dynamical frequency shifts $\delta \Omega_{\rm m}[y_0,\Omega]/2\pi$ from the three main sidebands. In the article we report on their spatial and spectral dependencies, shown in Fig.\,7.

We adjust the experimental results with an expression built from  eq. (\ref{eq.dWm}). We use the same dynamical temperature profile within the nanowire $\delta T[y,y_0,\Omega]$, where $y_0$ is the pump laser position, $y$  the vertical coordinate, as the one extracted in the deformation analysis conducted on the same nanowire (shown in Fig.\,4 of the manuscript).
In Figure 7d, we investigate the spectral response of the dynamical frequency modulation, at three different positions along the nanowire. To get rid of absorption inhomogeneities, it is convenient to analyse the normalized spectral response $\delta \Omega_{\rm m}[y_0,\Omega]/\delta \Omega_{\rm m}[y_0,0]$, and the dephasing of $\delta \Omega_{\rm m}[y_0,\Omega]$ which is independent from the absorption magnitude.   The different grey curves correspond to the normalized amplitude and argument of 
\begin{widetext}
\begin{equation}
\frac{\delta\Omega_{\rm m}[y_0,\Omega]}{\Omega_{\rm m}}
= 
  \frac{2}{L} \int_0^L{\left( \alpha_E \Sigma_m^2(y) +\alpha_L 
  \left( 4 \Sigma_m^2(y) + U_m^2(y)-1\right)
  \right)
  \delta T[y,y_0,\Omega]\,dy} \ \ + \  \mu\, \alpha_E^W  \Sigma_m^2(0)  \delta T[0^-,y_0,\Omega]\\
  \label{eq.dWmstrain}
\end{equation} 
\end{widetext}
by varying $y_0$ from 0 to L in $5\%$ steps. The last term corresponds to the contribution of the clamping area: similarly to the thermal wave which penetrates inside the tip due to the non ideal thermal contact, the deformation of the nanowire also put the tungsten tip into motion, causing a dynamical strain modulation inside the support \cite{WilsonRae2008,Anetsberger2008}. As a consequence, the thermal modulation in the tip will also impact the nanowire frequency. The prefactor $\mu=0.39$ is quasi unitary and is the same as the one employed in the spatial analysis of dynamical frequency modulation (Fig.\,7e). $\alpha_E^W= 4.8 \alpha_E$ is the temperature coefficient of the tungsten Young modulus. Despite this simple implementation, this added term allows to patch our model and to phenomenologically account for the dynamical strain located inside the nanowire support.
With those ingredients, the agreement with the response measurements is pretty good at low frequencies, when the strain contributions dominates, and very good at almost all frequencies for the blue curve taken at $y_0=0.27\,L$, again because at that location, the strain contribution dominates the global response. The deviation obtained at larger frequencies for the measurements conducted further away from the clamping area is due to the force gradients induced by the pump laser, which  modify the vibration frequencies of the nanowire \cite{Mercier2017}. \\
In the 1D case, a force gradient applied at the extremity of the nanowire will add its stiffness to the nanowire and cause a relative frequency shift given by 
$\frac{\delta \Omega_{\rm m}}{\Omega_{\rm m}} =\frac{-1}{2M_{\rm eff} \Omega_{\rm m}^2}\nabla F $. When the laser is positioned at a different position along the nanowire the force gradient effect will be reduced, and when the force presents a delay with respect to the light intensity modulation, the additional relative dynamical frequency shift will follow:
 
$$\frac{\delta \Omega_{\rm m}[\Omega]}{\Omega_{\rm m}} =-\frac{1}{2 M_{\rm eff} \Omega_{\rm m}^2}\nabla F [\Omega] \frac{u_m^2(y_0)}{u_m^2(L)} .$$

In the Figure 7e of the manuscript, we showed the spatial maps of the dynamical frequency shifts measured at different modulation frequencies $\Omega/2\pi$ with a constant modulation depth. The theoretical curves shown in gray are the ones obtained using  equation (\ref{eq.dWmstrain}). The colored curves are obtained by adding the force gradient contributions, given by the above expression, including both an instantaneous contribution from optical forces $\nabla F_{\rm opt}$ and a delayed contribution from the cross section modulation force $\nabla F_{\rm cs}[\Omega]=\nabla F_{\rm cs}\frac{\delta T[\Omega,y_0,y_0]}{\delta T [0,y_0,y_0]} $. Here we employed two fixed pre-factors all along the nanowire while a more complete analysis would  require analyzing  the different force contributions at each vertical position. Meanwhile, the agreement with the data, especially on the phase response which is less affected by absorption inhomogeneities,  is pretty good especially close to the clamping point, even without accounting for the force gradient contributions.

\begin{figure*}[t!]
\begin{center}
\includegraphics[width=0.85\linewidth]{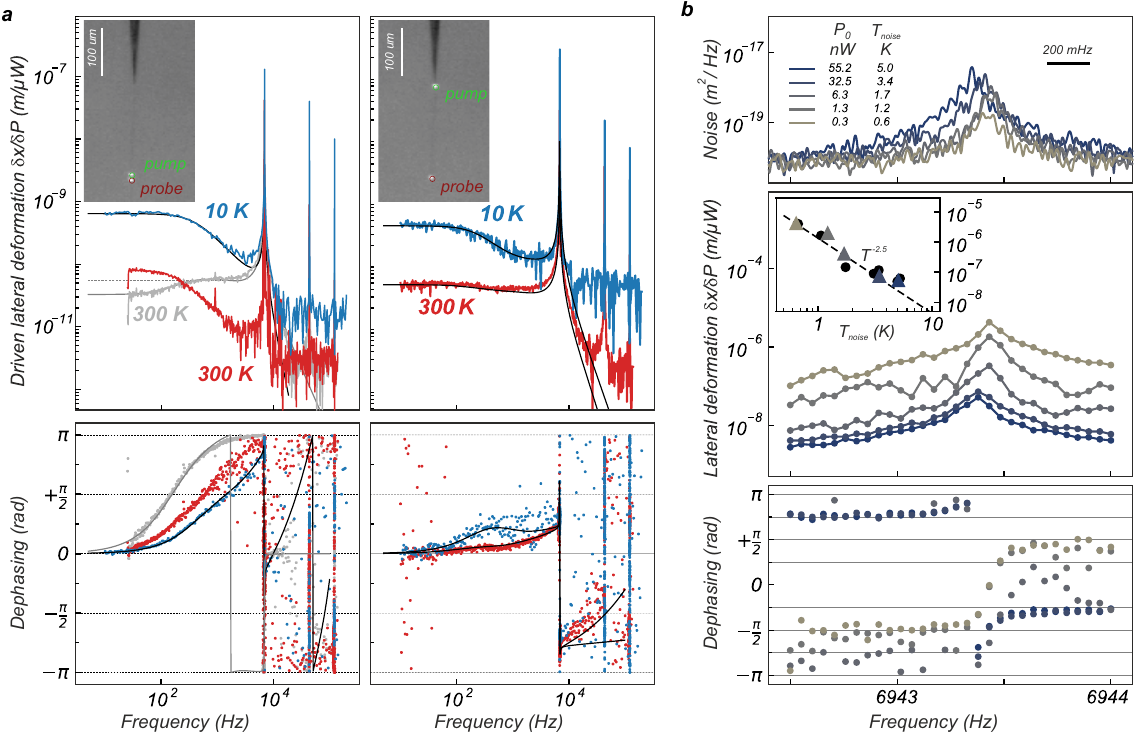}
\caption{
\textbf{Evolution of the photothermal response at cryogenic temperatures} ( $200\,\rm \mu m$ long, $180\,\rm nm $ wide nanowire) {\bf a)} Broadband  lateral responses measured when the pump laser is positioned at the vibrating (left) or clamped  (right) extremities of the nanowire, at noise temperatures of 10 K (blue)  and 300 K (gray and red after the subtraction of optical force contributions). All measurements are well adjusted, using an optical force  and a bilayer effect. At room temperature we use a diffusion coefficient $D = 2.2\times 10^{-5}\,\rm m^2/s $ and contact resistance $R_c$ below 7\,K/$\mu$W. At low temperature, the thermal actuation is dominant and is well fitted by two sets of parameters (D,R$_c$) whether the pump is at the tip ($D\simeq 2\times 10^{-4}$\,m$^2$.s$^{-1}$ and $R_c\simeq 28$\,K/$\mu$W) or close to the clamping area ($D\simeq 8.5\times 10^{-5} \,\rm m^2/s$ and $R_c\simeq 46\,\rm  K/\mu W$). The position dependence ought to be related to the non uniform temperature profile established along the nanowire combined with the temperature dependence of heat conductance and capacity of our SiC-nanowires. {\bf b)} Resonantly driven response of one mechanical mode of the nanowire realized at different average pump optical power (which sets a different static temperature profile along the nanowire). The noise temperature of the nano-oscillator  is deduced from its thermal noise spectra in the absence of optical modulation (top panel). The inset presents the evolution of the resonant displacement per $\mu$W of modulated optical power, with the noise temperature of the nanowire. A strong non linearity is observed: the dynamical response per input power modulation is 100 times stronger at low mean optical power, while the dephasing  also strongly evolve.
}
\label{fig-S4}
\end{center}
\end{figure*}

\section{Evolution at low temperatures}

We briefly report here some of the results we obtained at lower temperatures, first around 10\,K, then below. Those measurements were realized in the dilution fridge described in \cite{Fogliano2021a}.
The probe laser optical power is set around 1\,pW (633\,nm), so that it generates a negligible temperature increase ($<100\,\rm mK$) in  comparison to the temperature increase produced by the green laser, which was measured up to 10K in the  left panel. Those powers are however significantly smaller than the ones employed at room temperature.
On the same nanowire, we compared the photothermal responses obtained at room temperature and 10\,K (noise temperature of the nanowire), as shown in Fig.\,\ref{fig-S4}a, when positioning the pump laser at both extremities of the nanowire, to analyze the evolution of its contact resistance.  We first observe a drastic increase in the photothermal actuation efficiency, even exceeding the radiation pressure force  (materialized by a dashed gray line in the amplitude plot) when driving the nanowire at resonance.

At first glance, the broadband responses taken at low temperature present a similar shape  as in room temperature, suggesting that at 10\,K  the heat propagation inside the nanowire is apparently still in the diffusive regime. This is in accordance with previous results where we measured and analyzed the optical heating rates of the nanowire \cite{Fogliano2021a}.
Although one observes an increase of the resonant response due to the improved quality factor at reduced temperature \cite{Fogliano2021a}, the radiation pressure is no longer the dominant driving mechanism, even when pumping  the nanowire at its vibrating extremity (left plots) and the  response phase is not pinned to $0$ or $\pi$. In both configurations (pumping at the tip and at the clamping), we observe  an increased low frequency response (possibly due to a lower heat conduction coefficient $\kappa$), and a slightly faster thermal response, associated to an increase of the apparent heat diffusion coefficient.
It could be explained by a stronger reduction of the heat capacity compared to the heat conductivity, which is expected from bulk material properties.
The acceleration of the response is nonetheless lower than expected from tabulated data, most likely due to a consequent increase of the thermal contact resistance $R_c$. This is especially visible in the response phase plot when pumping close to the clamping point. They present an intermediary plateau, contrary to the signal at room temperature indicating a strong contact resistance (if absent, the lateral deformation should be almost flat in this frequency range).
The best fit obtained corresponds to a high value of $R_c$ with respect to the nanowire thermal resistance, $R_c/R_{\rm nw}\simeq 0.45$, which is very large for such a long nanowire.\\

We investigated the photothermal response of the nanowire at lower temperatures,  making use of the increased quality factor (Q $\simeq 10^5$) and of its resonant response to lower the pump power employed (see \ref{fig-S4}b), which allows to reach sub-K noise temperatures.
Here we focused on the higher frequency mechanical mode of the first longitudinal family.
The noise spectra measured in absence of pump modulation are used to determine the nanowire noise temperature, and the response measurements are shown for different mean optical powers (varying over two decades).
The  amplitude plots shows a strong  dependence to the mean optical pump power (x100 in the resonant responsivity) and thus on the noise temperature of the nanowire.
This non-linear behavior indicates that the correct understanding of those responses requires to revise the thermal model, and consider thermal propagation in a material whose thermal coefficients are no longer homogeneous all along the nanowire due to their temperature dependence.
The evolution of the  phase response at resonance with the mean optical power  also indicates a  change  in the dominant actuation mechanisms. The inset panel  shows the evolution of the resonant response amplitude with the noise temperature of the nanowire. featuring a rapid increase in the photothermal actuation efficiency at low temperatures, scaling as approx. $T^{-2.5}$.
It is this great variability observed at low temperatures that led us develop the room temperature experiment in order to better understand the different photothermal mechanisms at play in the system. This non-trivial behavior will be subjected to further investigation in order to disentangle the different mechanisms at play in the system, using the methods we have exposed in this work.


\end{document}